\documentclass[pdflatex,sn-nature]{sn-jnl}

\usepackage{amsmath,amssymb,amsfonts}
\usepackage{siunitx}

\usepackage{graphicx}
\usepackage{subcaption}

\usepackage{xcolor}

\hypersetup{
	colorlinks = true,
	urlcolor   = blue,
	linkcolor  = blue,
	citecolor  = red
}
\usepackage{cleveref}

\usepackage{acro}
\DeclareAcronym{2pa}{short=2PA,  long=two-photon absorption}
\DeclareAcronym{spad}{short=SPAD, long=single-photon avalanche diode}
\DeclareAcronym{pdf}{short=PDF,  long=probability density function}
\DeclareAcronym{pbs}{short=PBS,  long=polarising beam splitter}

\newcommand{\figroot}{figures}

\DeclareMathOperator*{\argmin}{arg\,min}

\begin{document}

\title[Event-by-event track imaging with an ultrafast plenoptic camera]{Event-by-event track imaging of charged particles with an ultrafast plenoptic camera
}

\author[1]{\fnm{Till} \sur{Dieminger}}
\author[2,3,4]{\fnm{Marco} \sur{Barbieri}}
\author[5,6]{\fnm{Vincenzo} \sur{Berardi}}
\author[1]{\fnm{Sa\'ul} \sur{Alonso-Monsalve}}
\author[7]{\fnm{Kodai} \sur{Kaneyasu}}
\author[7]{\fnm{Paul} \sur{Mos}}
\author[1]{\fnm{Tim} \sur{Weber}}
\author[1]{\fnm{Johannes} \sur{W\"uthrich}}
\equalcont{Now at the University of Zurich.}
\author[7]{\fnm{Claudio} \sur{Bruschini}}
\author[7]{\fnm{Edoardo} \sur{Charbon}}
\author*[1]{\fnm{Davide} \sur{Sgalaberna}}\email{davide.sgalaberna@cern.ch}

\affil[1]{\orgdiv{Institute for Particle Physics and Astrophysics}, \orgname{ETH Z\"urich}, \orgaddress{\street{Otto-Stern-Weg 5}, \postcode{8093} \city{Z\"urich}, \country{Switzerland}}}
\affil[2]{\orgdiv{Dipartimento di Scienze}, \orgname{Universit\`a degli Studi Roma Tre}, \orgaddress{\street{Via della Vasca Navale 84}, \postcode{00146} \city{Rome}, \country{Italy}}}
\affil[3]{\orgname{Istituto Nazionale di Ottica - CNR}, \orgaddress{\street{Largo E. Fermi 6}, \postcode{50125} \city{Florence}, \country{Italy}}}
\affil[4]{\orgname{INFN Sezione di Roma Tre}, \orgaddress{\street{Via della Vasca Navale 84}, \postcode{00146} \city{Rome}, \country{Italy}}}
\affil[5]{\orgdiv{Dipartimento Interateneo di Fisica ``Michelangelo Merlin''}, \orgname{Politecnico di Bari}, \orgaddress{\street{Via Orabona 4}, \postcode{70125} \city{Bari}, \country{Italy}}}
\affil[6]{\orgname{INFN Sezione di Bari}, \orgaddress{\street{Via Orabona 4}, \postcode{70125} \city{Bari}, \country{Italy}}}
\affil[7]{\orgdiv{Advanced Quantum Architecture Laboratory (AQUA)}, \orgname{\'Ecole polytechnique f\'ed\'erale de Lausanne (EPFL)}, \orgaddress{\street{Rue de la Maladi\`ere 71b}, \postcode{2000} \city{Neuch\^atel}, \country{Switzerland}}}

\abstract{
Modern elementary particle detectors often require
high-resolution three-dimensional particle tracking in large dense active volumes, which is difficult to achieve in standard devices without an extremely fine segmentation.
In recent work, a plenoptic camera instrumented with a single-photon avalanche-diode
array, called PLATON, demonstrated to image individual visible photons generated in monolithic scintillating material
and reconstruct their three-dimensional origin from a single-event exposure,
opening a route to
high-spatial-resolution detection of elementary particles in unsegmented scintillator.
However, those tests aimed to address photon sources of essentially zero geometric extent.
In this work we close the gap to the tracking regime.
First, we calibrated the PLATON prototype using light tracks in an organic plastic scintillator induced by a femtosecond infrared laser via two-photon absorption and characterised its capability to reconstruct photon-starved 3D lines with a novel algorithm,
achieving good agreement with custom Monte Carlo simulation.
Finally, the PLATON prototype was exposed to a test beam of pions at the CERN Proton Synchrotron.
For the first time, a plenoptic detector system was successfully used to detect and reconstruct the 3D tracks of elementary particles from 2D photon-starved images on an event-by-event basis marking a transformative approach to high-resolution measurements.
}

\maketitle

\section{Introduction}

Scintillator detectors have been central in major discoveries in particle physics \cite{Kolanoski:2020ksk,Rubbia_2022}
and are currently deployed at experiments detecting neutrinos
\cite{Hyper-Kamiokande:2025fci,Hyper-Kamiokande:2018ofw,Abi:2020qib,Abe:2019vii,NOvA:2022see,MINOS:2008hdf},
searching for
dark-matter candidates
\cite{dama-libra,doi:10.1126/sciadv.adv6503,ANDREEV2026170830,DarkSide-20k:2024yfq,Aprile:2020vtw},
or at colliders, such as the Large Hadron Collider (LHC) \cite{CMSHCAL:2007zcq,CMS:2009cmd}.
The ability to track particle trajectories in three dimensions (3D) is critical, especially in high-energy physics experiments, as it allows multiple particles generated in a single event to be identified and discriminated, and their kinematics to be inferred.
Such capability is normally achieved by segmenting the scintillator into thousands or millions of optically-separated voxels, whose spatial resolution is necessarily dictated by their size.
While this technology has demonstrated its robustness and reliability \cite{Sgalaberna:2017khy,T2K:2026zms,NOvA:2022see,MINOS:2008hdf,Joram:2015ymp,Papa:2023uqv}, scaling it up to large detectors while maintaining the same performance would be impractical and expensive, given the prohibitively large number of scintillating voxels and electronic readout channels.

The solution emerges from PLATON (PLenoptic imAge of Tracked photONs), reported in our previous work \cite{Dieminger2026}, a paradigm shift in particle detection: a system of ultrafast plenoptic cameras capable of operating in photon-starved regimes, to image particles propagating in a monolithic volume of scintillator.
A plenoptic camera, also known as light-field camera, combines a single main objective lens (main lens) with a micro-lens array (MLA) placed in front of an imaging photosensor \cite{Lippmann_1908_Epreuves_reversibles_donnant,Adelson_1992_Single_lens_stereo,2005_Ng_Handheld_lightfield,2006_Ng_Digital_Lightfield_Photography}.
Such a configuration is able to capture the ``light field'', which encodes the light intensity at each spatial point $(x,y,z)$ propagating in direction $(\phi,\theta)$~\cite{Levoy_1996_Light_field_rendering}.
Effectively, each microlens, together with the subset of sensor pixels behind it, acts as a tiny camera viewing the intermediate image from a slightly different viewpoint.
As a result, the parallax between neighbouring micro-images encodes depth, which in the light-intensity regime can range from micrometres to several metres, depending on the optics design.
The PLATON plenoptic camera is equipped with a single-photon avalanche diode (SPAD) array imaging sensor,
whose fast response allows short-duration sub-\si{\micro\second} frames to be recorded, drastically rejecting most of the random-noise background by coincidence and thus reducing ambiguities in the post-processed 3D image.
As a result, PLATON leverages backward ray tracing of single photons from the pixels of the SPAD array to the scintillator and determines their origins by interpolation.
Finally, a system of multiple plenoptic cameras viewing the same scintillator volume from orthogonal directions would achieve the same resolution in depth as laterally, thus of a few hundred micrometres.
In fact, the depth measurement of one camera would be replaced by the lateral measurement of the cameras matched on the orthogonal view.

The first PLATON prototype, equipped with SwissSPAD2 \cite{Ulku2019},
succeeded in reconstructing the 3D position of a light point with $\mathcal{O}$(mm) depth resolution and sub-millimetre lateral resolution over a depth of 10 cm, even with fewer than 20 detected photons.
The gate window of a single frame, tunable between 10 \si{\nano\second} and 100 \si{\micro\second}, allowed to maximise the signal-to-noise ratio.
Single positrons from a $^{90}$Sr source were detected and localised in a monolithic plastic scintillator block \cite{Dieminger2026}.
Moreover, optical simulations showed the potential of a system of PLATON cameras to fully reconstruct the 3D tracks of multiple particles, including protons with momenta down to 200 MeV/c, produced in GeV neutrino interactions.
In a $10 \times 10 \times 10$ cm$^3$ scintillator,
a 3D spatial resolution
of 200 \si{\micro\metre}
was obtained using a transformer model.
Moreover, preliminary simulation studies with point-like light sources hint at a spatial resolution between 1 and 3 millimetres in a 1 m$^3$ scintillator, even though the optics was not optimised for that volume size \cite{Dieminger2026}.

However, a fundamental step in the full validation of the PLATON concept was still missing: neither light tracks nor particle trajectories had been reconstructed experimentally with the PLATON prototype.
In this work, we fill this gap by demonstrating experimentally that the PLATON detector fulfils the requirements for tracking ionising particles in a scintillator volume on an event-by-event basis in photon-starved conditions.
To this aim, the original ray tracing algorithm had to be extended for particle tracking.
Non-linear optical effects occurring in the scintillator material~\cite{GoeppertMayer1931,Denk1990,Auffray2015} have been harnessed in order to accurately produce test tracks by means of laser two-photon absorption
(2PA):
the simultaneous absorption of two photons at 800 nm around its focal region in the scintillator induces a localised, isotropically emitted light, similar to that produced by a traversing charged particle.
The characterisation of the PLATON prototype with
2PA
in a fully controlled environment made it possible to employ
it
in a beam test at CERN to detect and reconstruct the 3D track of \SI{15}{\giga\electronvolt\per c} pions from 2D photon-starved images.
To the best of our knowledge, this is the first time that a plenoptic system has been used to track elementary particles interacting in a monolithic scintillator volume in 3D on an event-by-event basis. Our results open a path towards a new generation of particle tracking detectors, which may also find applications outside fundamental research, such as medical imaging.

\section{Results}
\label{sec:results}

\subsection{Reconstruction of light tracks}
\label{sec:reconstruction}

A plenoptic camera used under bright illumination records a small sub-image for each of its microlenses.
Neighbouring microlenses share features which can be identified and employed to retrieve the parallax, eventually yielding the depth of objects from a 2D picture by standard light-field algorithms~\cite{Perwass2012,Adelson1992,Jeon2015,Ng2005}. In the photon-starved regime typical of particle detection one cannot rely on the post-processing of individual frames, as these normally contain too few photons -- of the order of a few tens -- to form the well-structured sub-images the traditional algorithms require.

This important difference calls for a different approach to retrieving the source of the photons. We use the chief-ray method introduced in~\cite{Dieminger2026}. Such rays are determined by connecting each active pixel to the centre of the corresponding microlens and then back-propagated to the scintillator, treating the main lens as a thick lens and accounting for the air-scintillator interface. The intersection of the resulting rays in the scintillator volume identifies the origin of the scintillation emission.

\begin{figure}
	\centering
	\begin{subfigure}[b]{0.45\linewidth}\centering\includegraphics[width=\linewidth]{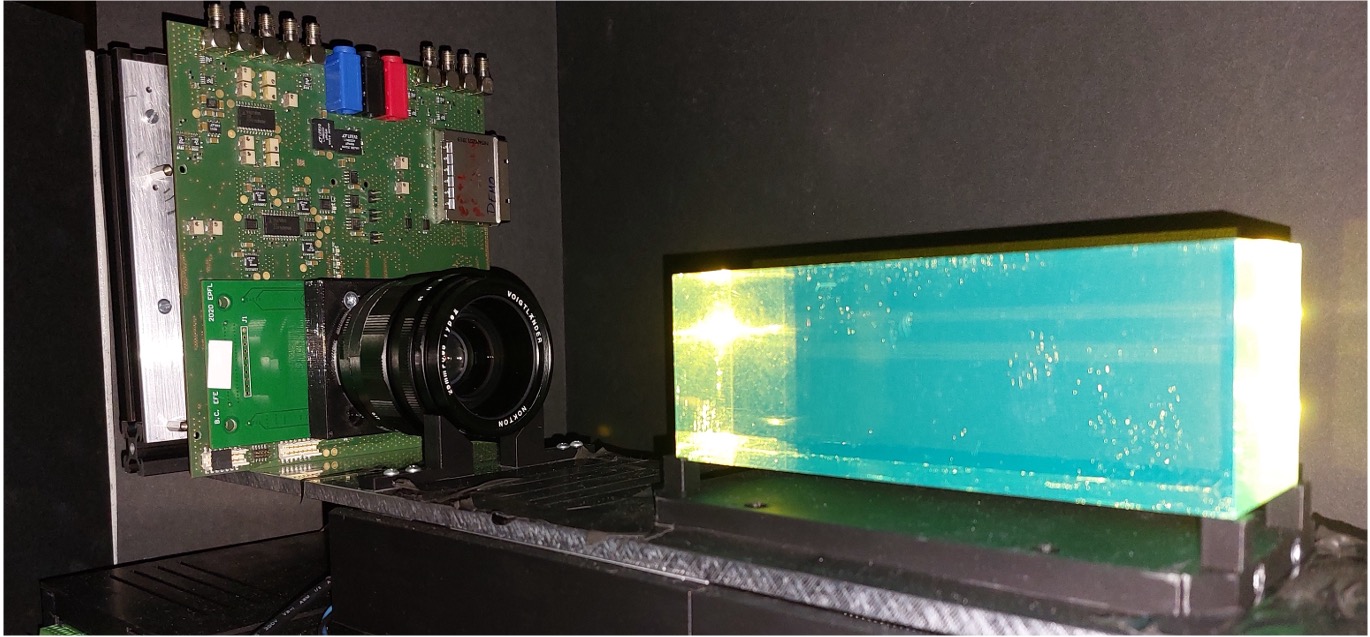}\caption{}\label{sfig:method-photo}\end{subfigure}\hfill
	\begin{subfigure}[b]{0.22\linewidth}\centering\includegraphics[width=\linewidth]{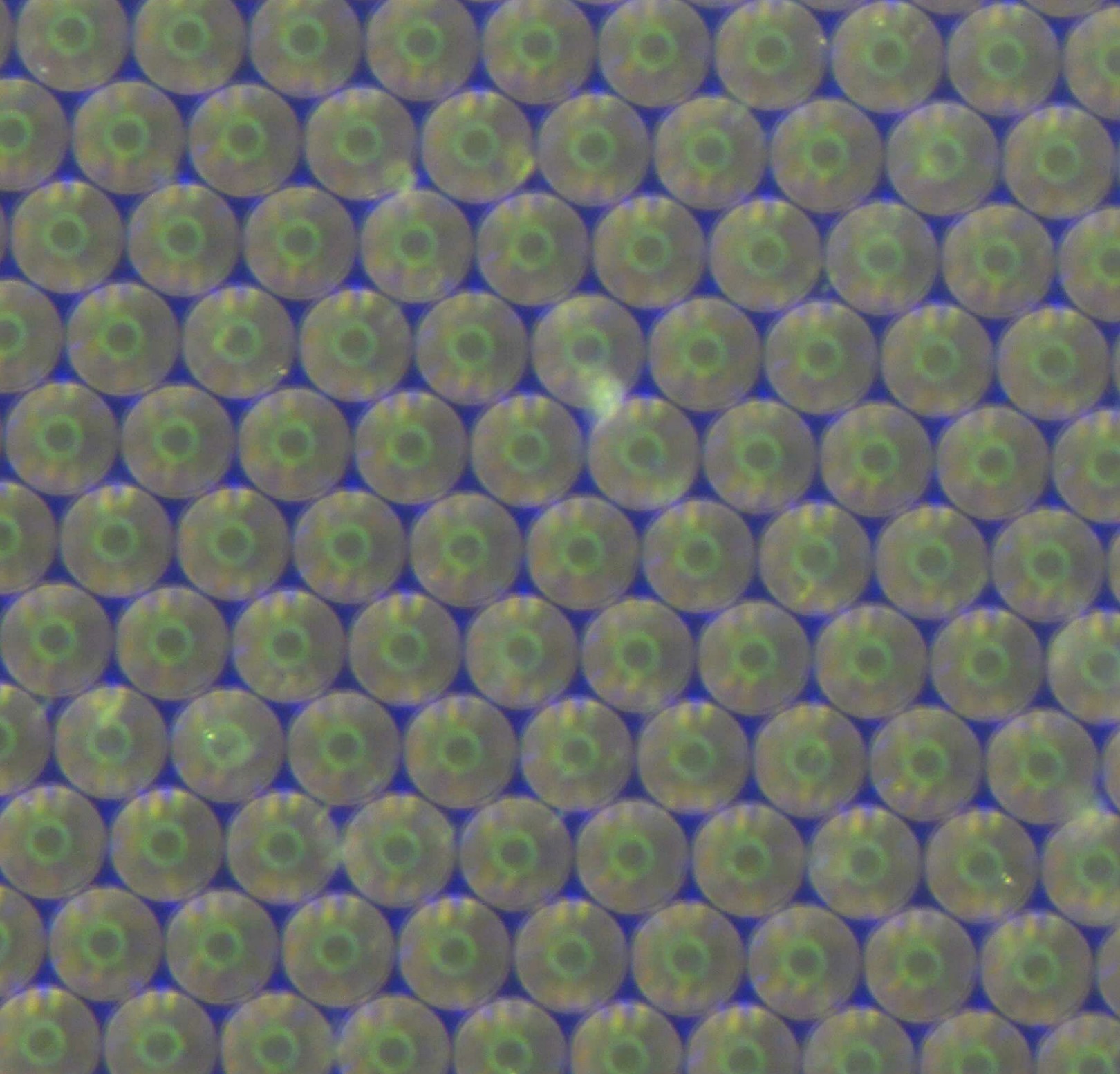}\caption{}\label{sfig:microlenses}\end{subfigure}\hfill
	\begin{subfigure}[b]{0.2\linewidth}\centering\includegraphics[width=\linewidth]{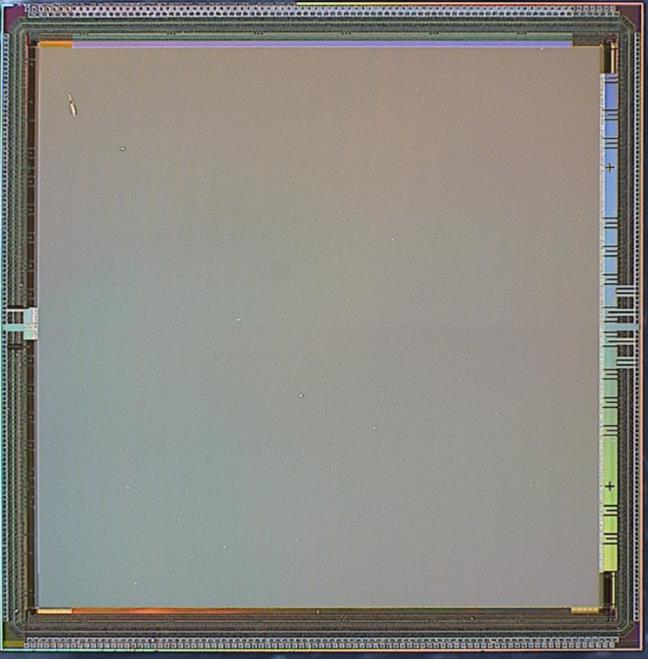}\caption{}\label{sfig:method-ss2}\end{subfigure}\\[1.5ex]
	\begin{subfigure}[b]{\linewidth}\centering\includegraphics[width=0.75\linewidth]{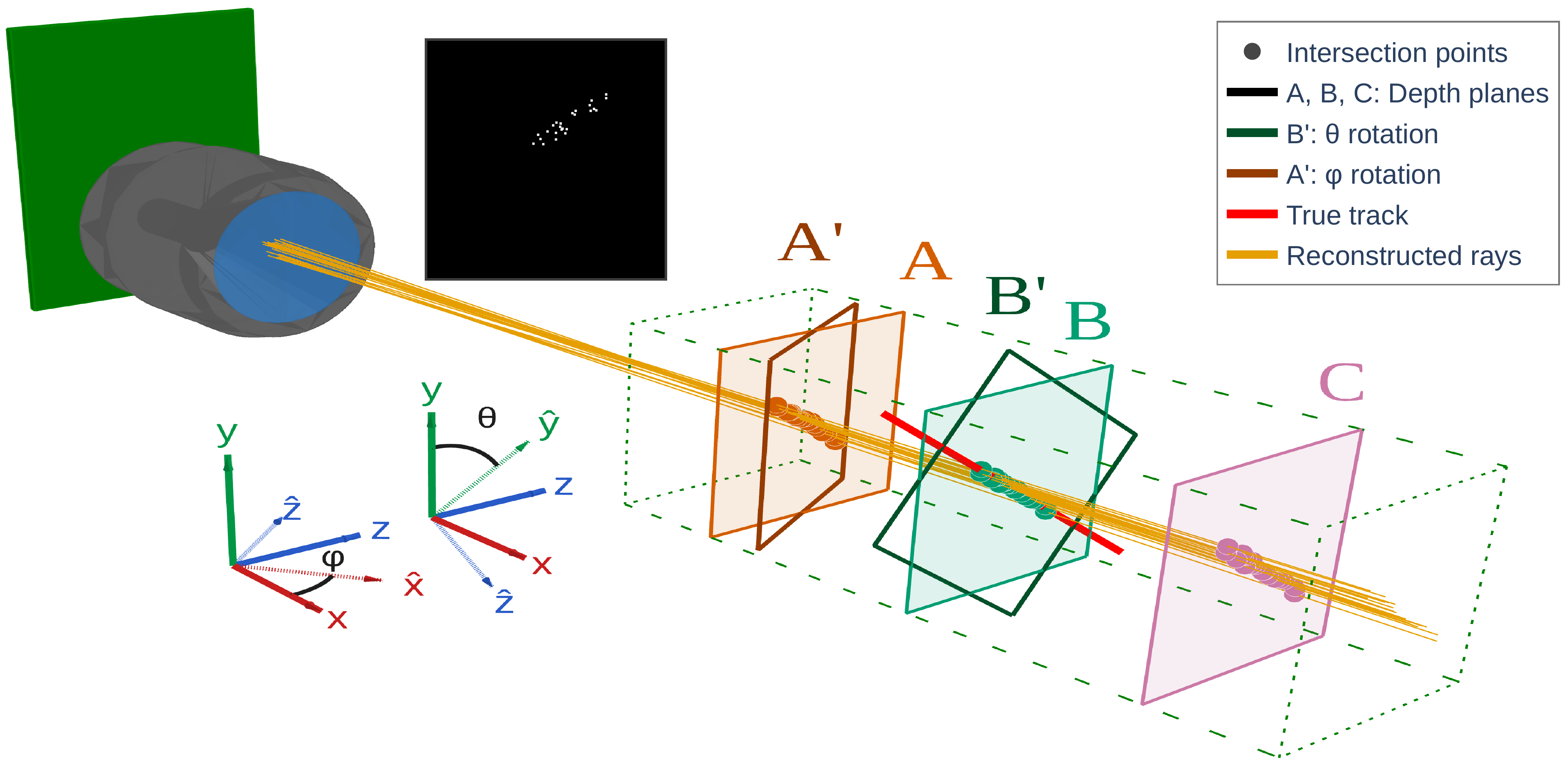}\caption{}\label{sfig:method-3d}\end{subfigure}
    \vspace{1.5ex}
	\begin{minipage}[c]{0.46\linewidth}
        \centering
        \begin{subfigure}[c]{\linewidth}\centering\includegraphics[width=\linewidth]{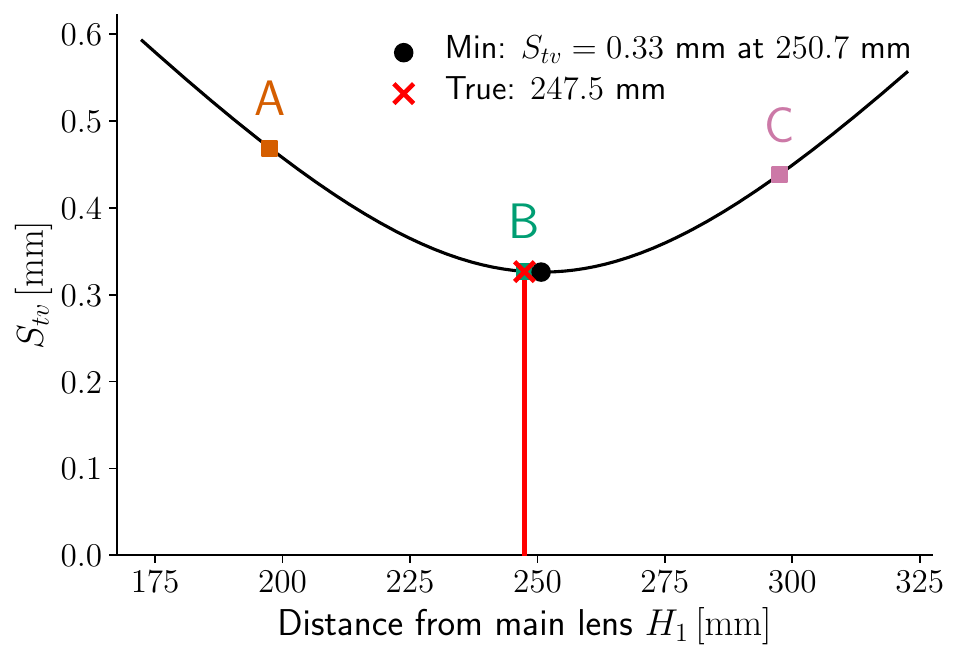}\caption{}\label{sfig:method-sigma}\end{subfigure}
    \end{minipage}
    \begin{minipage}[c]{0.50\linewidth}
        \centering
        \begin{subfigure}[t]{\linewidth}\centering
        \includegraphics[width=0.42\linewidth]{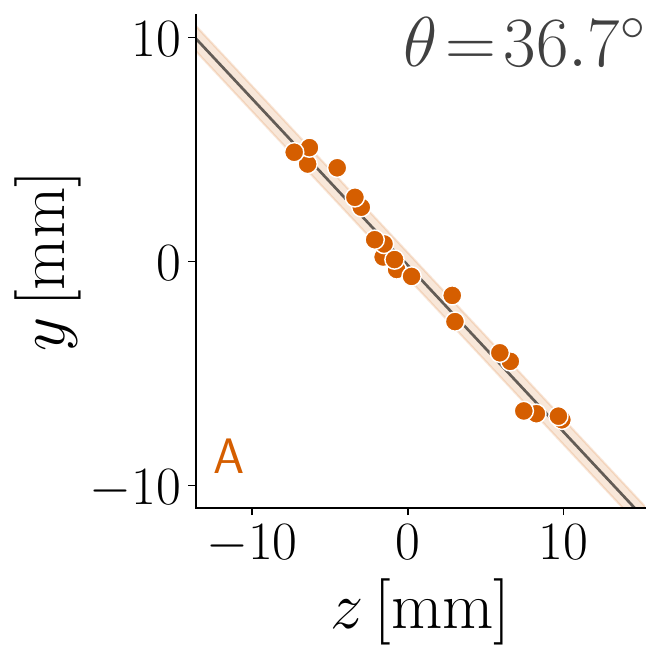}
        \hspace{1mm}
        \includegraphics[width=0.42\linewidth]{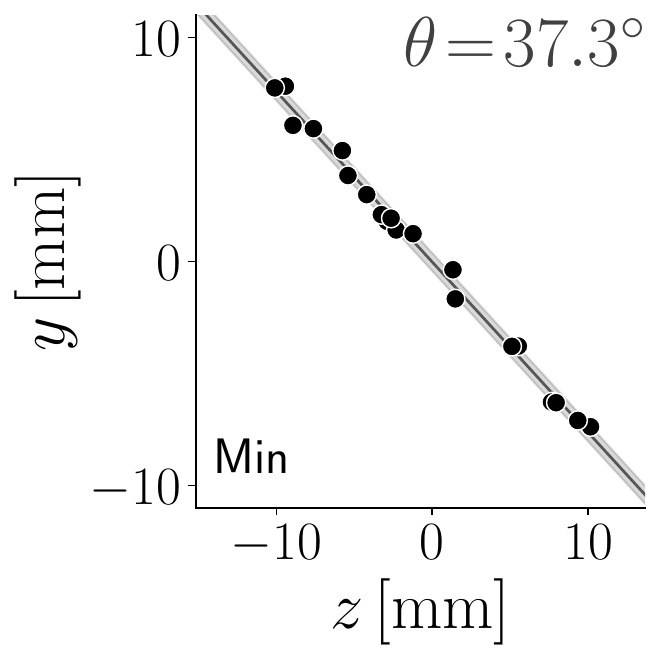}\\[0.5ex]
        \includegraphics[width=0.42\linewidth]{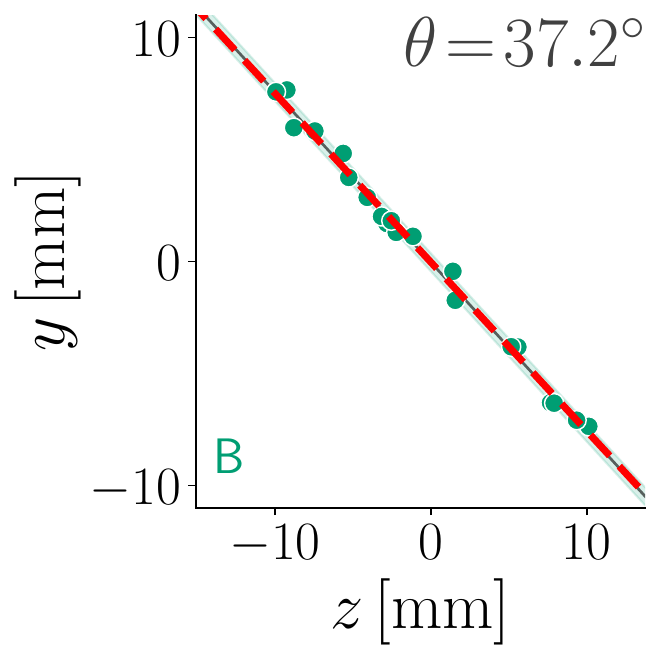}
        \hspace{1mm}
        \includegraphics[width=0.42\linewidth]{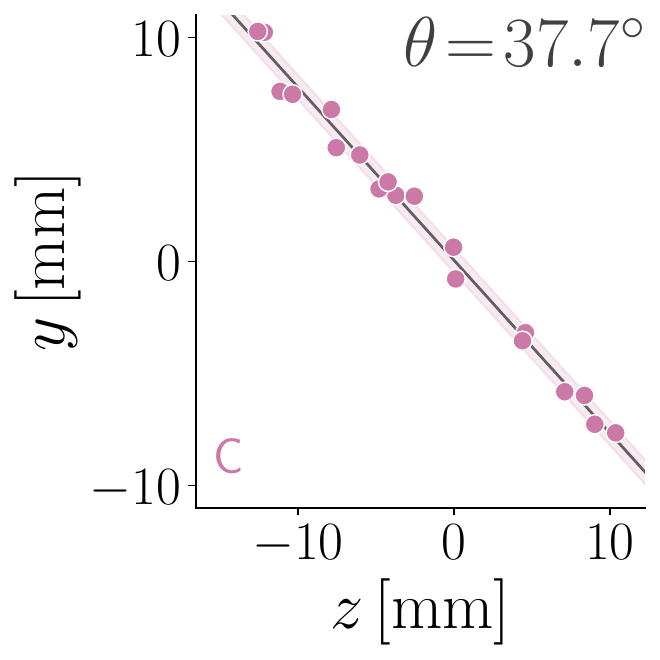}
        \caption{}\label{sfig:method-depthscan_intersection}
        \end{subfigure}
    \end{minipage}\hfill
    \caption{
    The PLATON prototype and the track event reconstruction method.
	\subref{sfig:method-photo} Photograph of the PLATON detector prototype.
	\subref{sfig:microlenses} Photograph of the PLATON microlens array.
	\subref{sfig:method-ss2} Micrograph of the SwissSPAD2 sensor.
	\subref{sfig:method-3d} Reconstructed photon rays (orange) for a single simulated light track (red), intersected by transverse planes at three depths, labelled A, B and C; the green box marks the observation volume.
    The primed outlines show the rotations explored by the fit: A$'$ is plane A rotated by $\phi$, B$'$ is plane B rotated by $\theta$, and the two coordinate systems indicate the corresponding rotations.
    The inset shows the simulated SwissSPAD2 frame of the same event, from which the rays are reconstructed.
    \subref{sfig:method-sigma} Transverse spread $S_{tv}$ as a function of the distance from the principal plane $H_1$ of the main lens. The spread passes through a minimum, which determines the reconstructed depth.
    The red line marks the true depth of the object, and the squares labelled A, B and C correspond to the planes shown in \subref{sfig:method-3d}.
    \subref{sfig:method-depthscan_intersection}
    Intersection points in the transverse ($y$--$z$) plane at the depths A, B and C, and at the depth of minimum spread (Min).
    }
    \label{fig:prototype}
    \label{fig:track-reco-method}
\end{figure}

In the original work \cite{Dieminger2026}, light point sources were successfully imaged with the chief-ray method.
However, the algorithm could not reconstruct any object geometry more complex than that of a single point.
In the present study, instead,
the object consists of a track along which photons are emitted at different points.
Thus, the rays emerge from an extended object, which can be approximated as a straight line.
Referring to the geometry in Fig.~\ref{fig:track-reco-method}, the propagation and intersection of the chief rays with a plane at depth $d$ results in a set of aligned points and thus a track.
The rotations of the plane at depth $d$ by an angle $\theta$ around the normal axis, coincident with the camera optical one, and by $\phi$ about the in-plane axis parallel to $y$,
define the track direction, as depicted in \cref{fig:track-reco-method}.
Thus, the best-fit plane is obtained by finding the set of \{$d,\theta,\phi$\} values that minimise $S_{tv}$, i.e. the transverse spread of the track.
It is worth noting that, since the depth resolution of the camera is about \SI{20}{\milli\metre} for point sources with 5--10 detected photons, its sensitivity to a change in $\phi$ is rather limited,
especially if the visible length of the tracks is short, as is the case here.
Thus, for simplicity, in the reconstruction we fix $\phi=0$, assuming that all the tracks lie in a plane parallel to the camera optical axis.
Details of the reconstruction method can be found in \cref{sec:fitting}.

\subsection{Characterisation and reconstruction with laser-induced two photon-absorption}
\label{sec:2pa-reco}

To ensure the reliable detection of beam particles, it is first crucial to characterise the detector in a controlled manner using light-track objects.
Two-photon absorption (2PA) in the scintillator block induced by a focused femtosecond infrared laser beam produces a short, straight scintillation light track, mimicking the ionisation of a through-going charged particle, though at a known and reproducible position and orientation, providing an end-to-end ground truth.
The scintillator volume was scanned by generating 2PA-induced light tracks in depth along the optical axis with a motorised translation stage, while data were acquired with the PLATON camera.
The scintillator under test is a \qtyproduct{50 x 50 x 150}{\milli\metre} block of EJ-262, polished on all six faces by diamond milling \cite{datasheet_EJ_260_262} held in an aluminium frame.
The 2PA optical layout and acquisition setup are detailed in \cref{sec:setup}.
To measure the noise pattern of the photosensor, a background-only data sample was collected while the laser was turned off.
This allowed us to map and discard the 5\% noisiest pixels (screamers) in the image analysis.

The acquired datasets can be manipulated to test the reconstruction performance at different illumination and noise levels, from light intensity to photon-starved images.

\subsubsection{Reconstruction of light intensity images and calibration}
\label{sec:2pa-reco-intensity}

Light intensity images are formed by adding up \num{163840} acquired frames, each corresponding to an exposure of \SI{100}{\nano\second}.
Once the image is formed, comparison with our custom optical simulation (see details in \cite{Dieminger2026}) allows the contributions from random noise, residual environmental light and, to a lesser extent, light in the tail of the 2PA process to be identified with a high level of confidence.
Pixels with a recorded intensity below \SI{10}{\percent} of the most illuminated pixels were omitted from the image analysis in order to obtain a thin light track (thresholding).
Example images can be found in \cref{fig:setup}.
By scanning the scintillator block between the two extremes, the distribution of the light track among the microlens sub-images changes visibly.
The intensity images obtained by positioning the 2PA track at the front and the rear of the scintillator block show clear differences:
the closer the track is to the camera, the wider the vertical spread of microlenses containing a sub-image.
This feature is used in classical reconstruction algorithms to determine the depth of an object imaged by a plenoptic camera.

In this work, the track reconstruction performance is evaluated in terms of depth and lateral resolutions, which we define as the half-width of the central 68\% one-dimensional interval between the true and reconstructed positions.
Both resolutions are evaluated for the track centroid, which provides a compact measure of the distance between the true and reconstructed tracks.
Along the $y$ axis, the vertical centroid gives the lateral resolution of the track.
Instead, along the $x$ axis, the depth resolution of the track centroid is determined
(see the definition of the coordinate system in \cref{fig:track-reco-method}).
It is worth noting that the 2PA light track was kept on the camera optical axis, because it could not be moved vertically along $y$ due to a limitation of the movement stage.

\begin{figure*}
	\centering
	\begin{subfigure}[t]{0.3\linewidth}\centering\includegraphics[width=\linewidth]{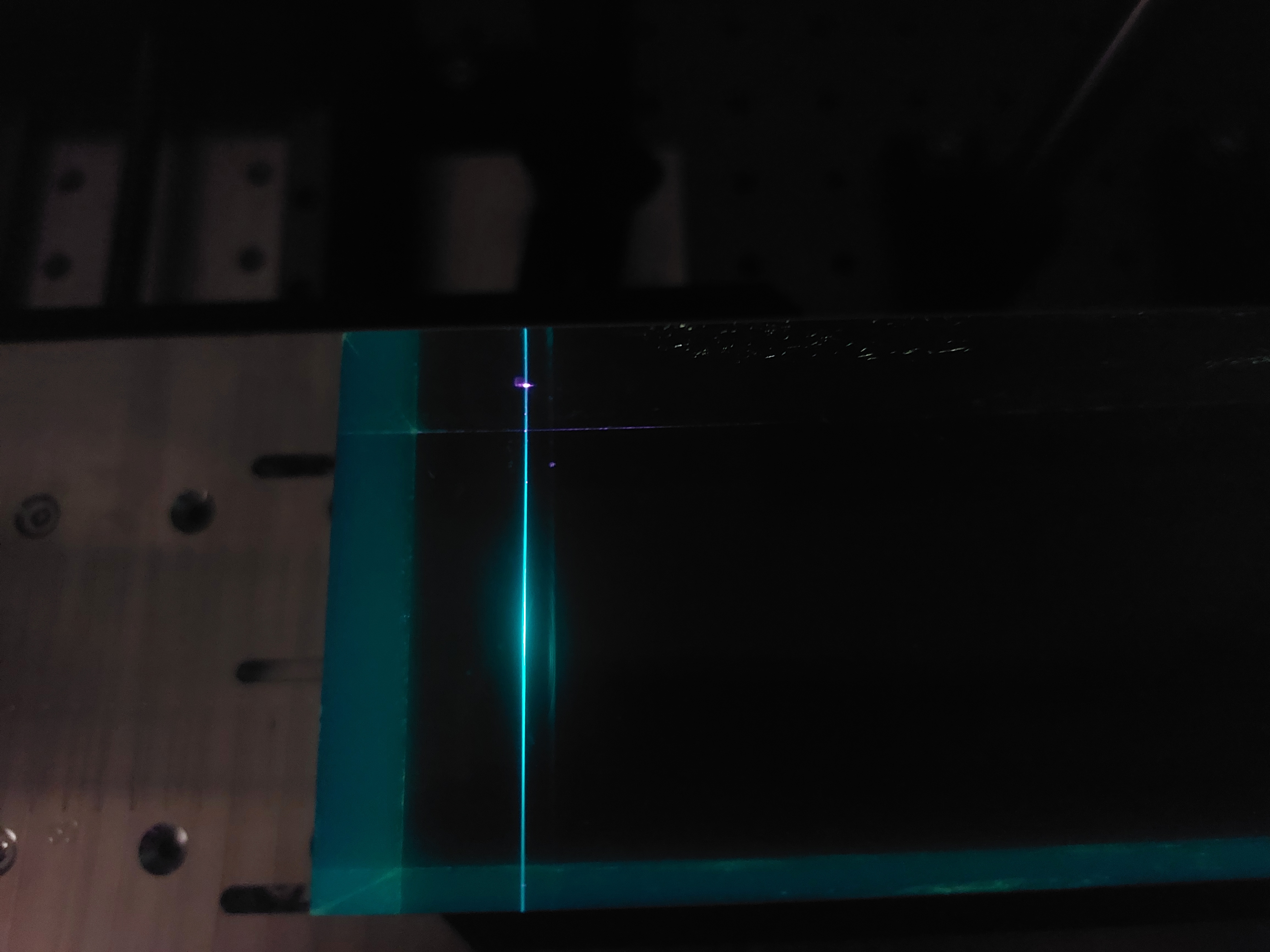}\caption{}\label{sfig:2pa-photo}\end{subfigure}\hfill
    \begin{subfigure}[t]{0.3\linewidth}\centering\includegraphics[width=\linewidth]{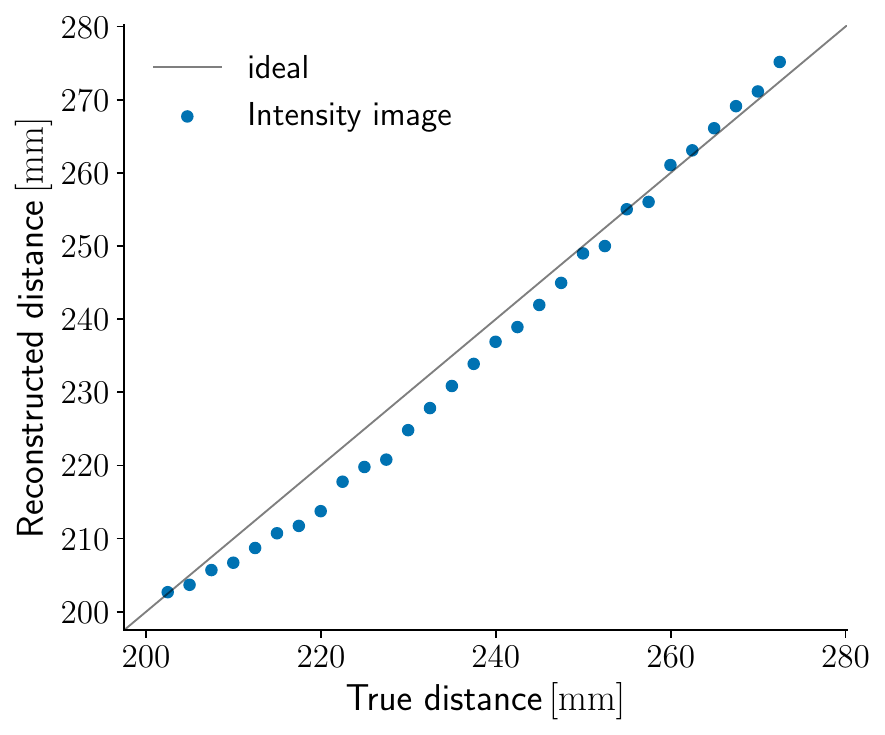}\caption{}\label{sfig:res-summed-depth}\end{subfigure}\hfill
	\begin{subfigure}[t]{0.32\linewidth}\centering\includegraphics[width=\linewidth]{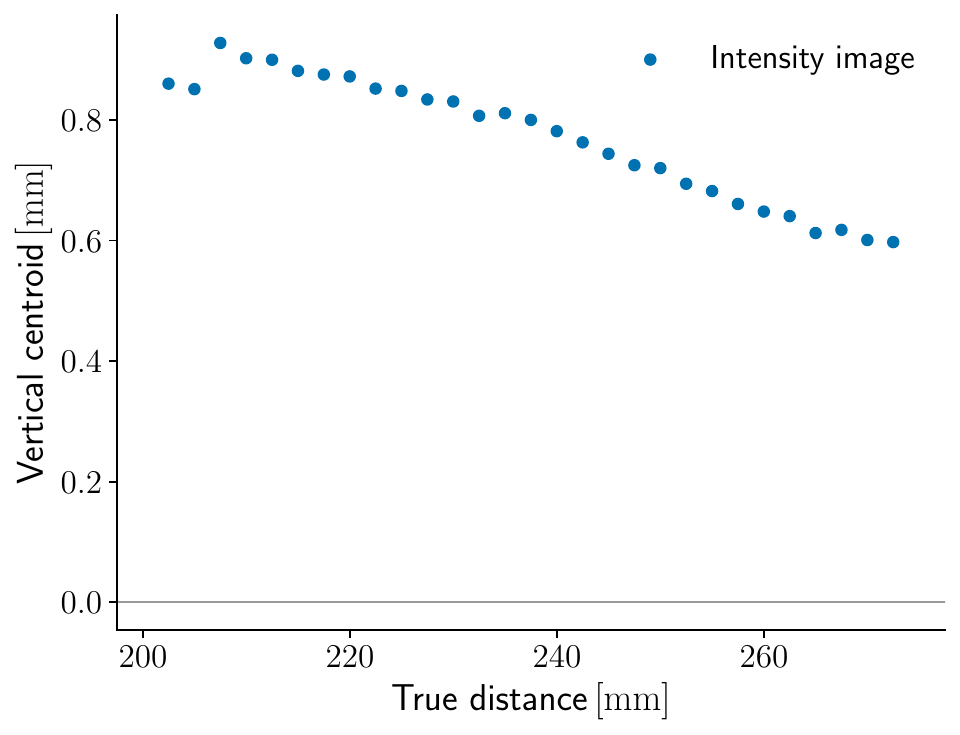}\caption{}\label{sfig:res-summed-lat}\end{subfigure}\hfill
	\begin{subfigure}[t]{0.32\linewidth}\centering\includegraphics[width=\linewidth]{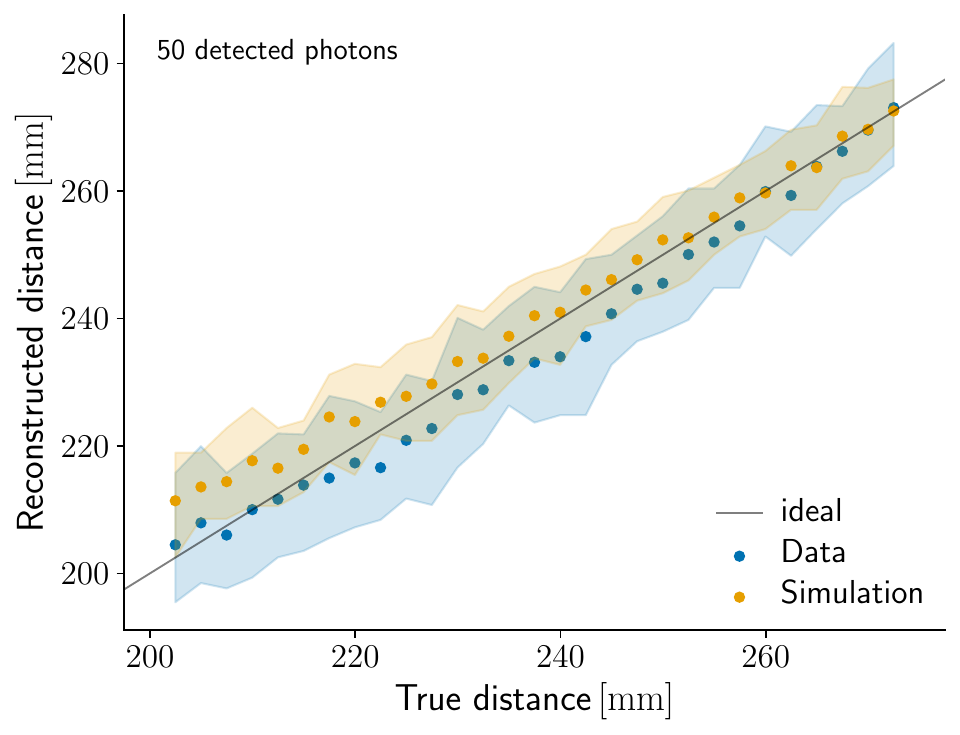}\caption{}\label{sfig:res-depth-n50}\end{subfigure}\hfill
	\begin{subfigure}[t]{0.32\linewidth}\centering\includegraphics[width=\linewidth]{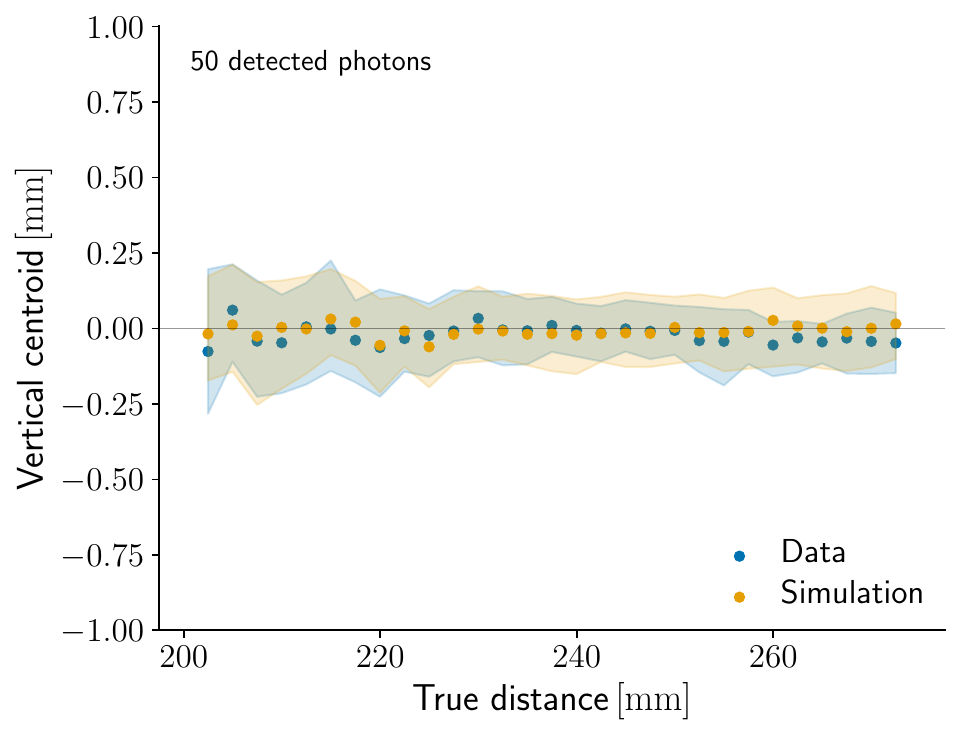}\caption{}\label{sfig:res-lat-n50}\end{subfigure}
	\begin{subfigure}[t]{0.32\linewidth}\centering\includegraphics[width=\linewidth]{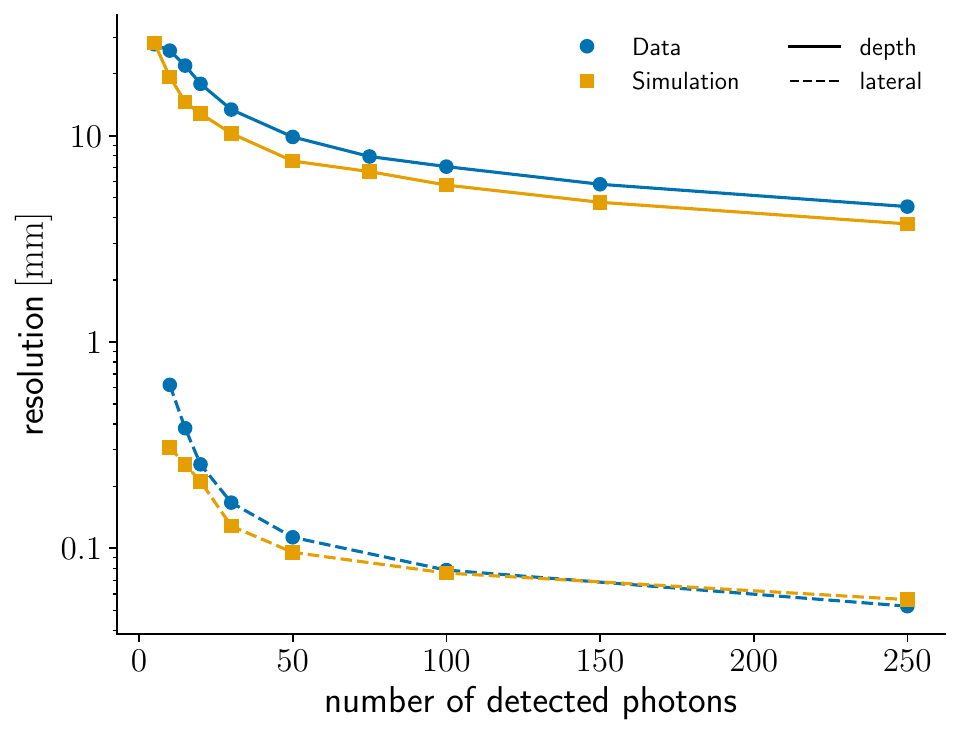}\caption{}\label{sfig:res-band}\end{subfigure} \\
	\begin{subfigure}[t]{0.4\linewidth}\centering\includegraphics[width=\linewidth]{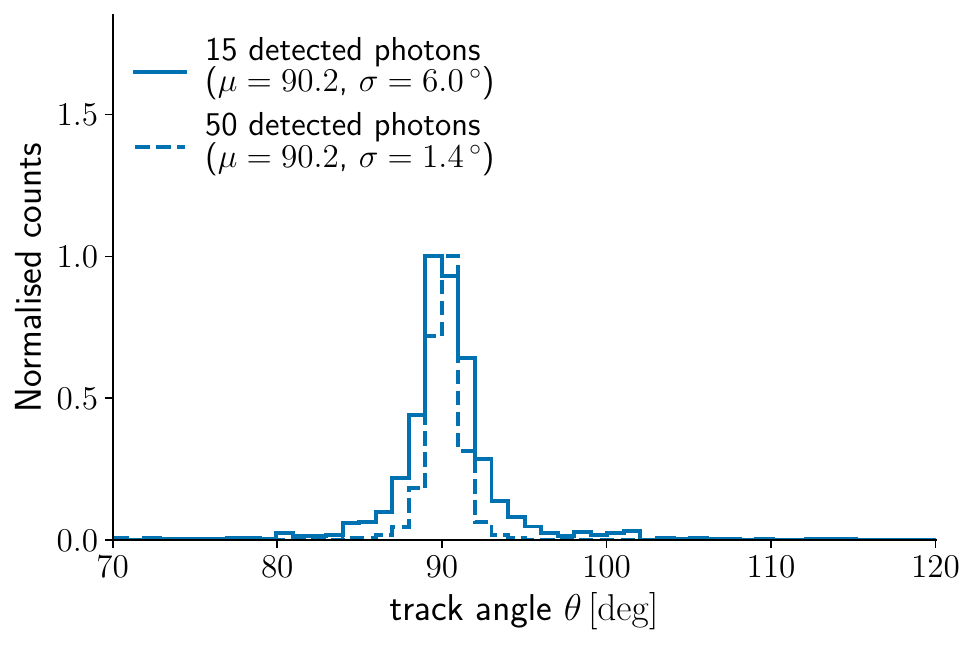}\caption{}\label{sfig:2pa-theta}\end{subfigure}%
	\begin{subfigure}[t]{0.4\linewidth}\centering\includegraphics[width=\linewidth]{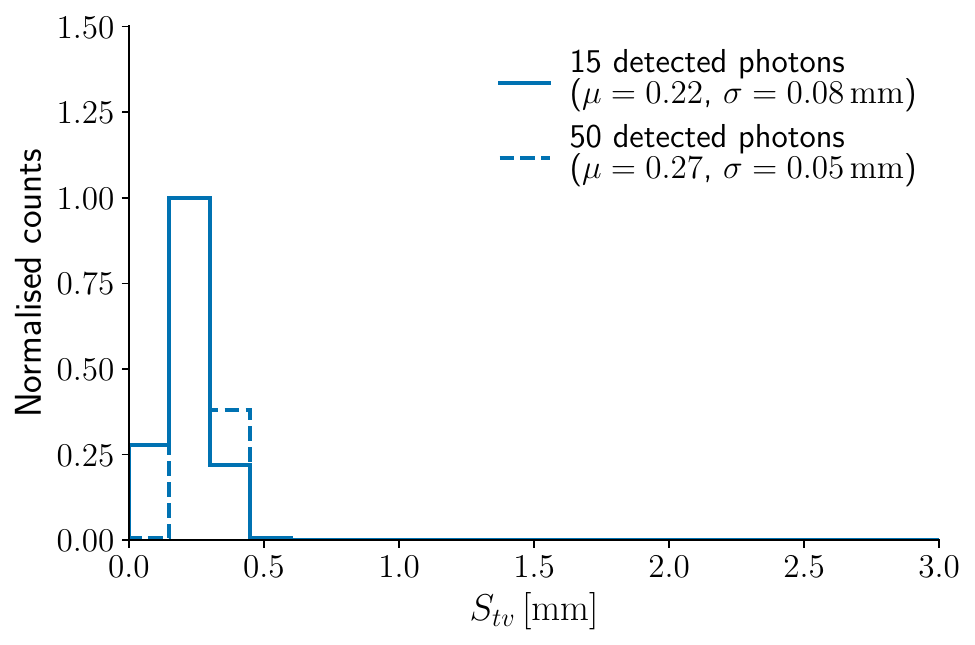}\caption{}\label{sfig:2pa-stv}\end{subfigure}
	\caption{
    Reconstruction of 2PA laser tracks at different depths in the scintillator block.
    \subref{sfig:2pa-photo} Photograph of the 2PA track in the scintillator.
    \subref{sfig:res-summed-depth} Reconstructed versus true depth centroid position.
    \subref{sfig:res-summed-lat} Vertical centroid with respect to the nominal position.
    Reconstructed versus true \subref{sfig:res-depth-n50} depth and \subref{sfig:res-lat-n50}
    vertical positions for frames with 50 detected photons sampled from PDFs.
    \subref{sfig:res-band} Depth (solid) and vertical (dashed) resolutions as a function of the number of detected photons for data (blue) and simulation (orange).
    Distribution of the reconstructed 2PA track
    \subref{sfig:2pa-theta} angle, $\theta$, and
    \subref{sfig:2pa-stv} transverse spread, $S_{tv}$, for frames with 15 (solid) and 50 (dashed) detected photons sampled from PDFs.
    }
   \label{fig:2pa}
	\label{fig:intensityReco}
	\label{fig:summedDepth}\label{fig:lateraldrift}%
	\label{fig:datamc}\label{fig:recoraw}\label{fig:rawband}
\end{figure*}

As shown in \cref{sfig:res-summed-depth}, the reconstructed 2PA track depth closely follows the true depth.
In \cref{sfig:res-summed-lat}, one can see a near-constant tendency of the vertical centroid position to drift by about 0.1 mm for a depth variation of 20 mm, i.e. $\sim$\SI{5}{\milli\radian}.
Moreover, an offset of $\sim$\SI{0.7}{\milli\metre} is also observed.
These effects were studied and reproduced with our custom optical simulation.
The light track is generated by uniformly distributing scintillation photons along a line for a given direction and depth in the scintillator volume.
Thus, the same reconstruction method as described in \cref{sec:reconstruction} can be applied.
First, we were able to reproduce the offset in simulations, which can be explained with the misalignment of the laser beam with respect to the scintillator,
to which our reconstruction method is sensitive.
The \SI{2}{\milli\metre} pinhole apertures used for the alignment cannot distinguish straight paths passing anywhere within their diameter, leaving a residual uncertainty.
Details of the laser alignment procedure can be found in \cref{sec:setup}.
The drift can be explained by a slight tilt of the translation stage with respect to the optical axis: if the movement axis is tilted by $\sim$\SI{5}{\milli\radian}, points close to the sensor sit above the optical axis while points farther away sit below it.
Thus, the relative position of the laser beam and translation stage with respect to the scintillator block and camera optical axis was calibrated by fitting the simulated position and orientation of the laser with respect to the scintillator to the data.

The depth and vertical-centroid resolution after calibration are shown in \cref{fig:datamc},
with the latter being around \SI{0.2}{\milli\metre}.
It is worth noting that the resolutions obtained from data and simulation are similar, indicating the quality of the custom optical simulation.
The inclusion of secondary 2PA effects, such as the non-Gaussian tail, can be studied without thresholding (see \cref{sec:2pa-reco-intensity}). Results in the Supplementary Material show that the vertical-centroid resolution worsens by about
\SI{0.1}{\milli\metre}.

\subsubsection{Reconstruction of photon-starved images}
\label{sec:2pa-reco-photon-starved}

Photon-starved frames are built from normalised light-intensity images used as \ac{pdf}: single images are randomly sampled for a specific number of detected photons, typically between ten and a few hundred.
PDFs are built after discarding the screamer pixels at different light-track depths both with and without noise.
In the latter case, they are obtained with thresholding,
as described in \cref{sec:2pa-reco-intensity}.

The 2PA track reconstruction was performed after the calibration as described in \cref{sec:2pa-reco-intensity}.
The correlation between the true and reconstructed depth or vertical-centroid position is shown in \cref{fig:datamc} for the \ac{pdf}-sampled frames.
While the depth resolution is about \SI{21.9}{\milli\metre} for 15 detected photons and \SI{9.8}{\milli\metre} for 50 detected photons, converging to \SI{4.5}{\milli\metre} above 200 photons,
the vertical centroid position easily reaches sub-millimetre resolution,
i.e. \SI{0.38}{\milli\metre} for 15 detected photons and \SI{0.11}{\milli\metre}
for 50 photons.

The mean transverse spread of the reconstructed tracks, $S_{tv}$ in \cref{sfig:2pa-stv}, quantifies the apparent thickness (blur) of the track.
It is \SI{0.22}{\milli\metre} for 15 and \SI{0.27}{\milli\metre} for 50 detected photons, converging to \SI{0.30}{\milli\metre} for a few hundred photons.
This increase indicates the need for further optimisation of the algorithm that rejects outlier rays, especially for large samples of detected photons.

Results were also obtained with frames generated from the PDFs without thresholding, i.e. not cleaned of either secondary signal components, such as reflections or tail of the 2PA process, or dark counts, and thus under non-ideal conditions.
For 50 detected photons, a depth resolution of \SI{17}{\milli\metre} is obtained.
The lateral resolution was only moderately affected, degrading to
\SI{0.24}{\milli\metre} for 50 detected photons.

Finally, we investigated the impact of dark counts in the absence of the 2PA-process tail. Dark counts are visible across the whole sensor surface, whereas the 2PA tail is observed in the sub-image of each microlens, which independently sees a different view of the same 2PA track.
Thus, while thresholding was applied, pixels with fewer than 3 counts in the PDF were retained.
With dark counts included but the 2PA-process tail excluded, the depth resolution becomes \SI{14.2}{\milli\metre} for 50 detected photons, with a lateral resolution of
\SI{0.17}{\milli\metre}.

\subsection{Pion test beam reconstruction}
\label{sec:test-beam-reconstruction}

After the characterisation with 2PA light tracks, the PLATON prototype was exposed to a beam of $\SI{15}{\giga\electronvolt\per c}$ negatively charged pions at the CERN Proton Synchrotron.
The setup is described in detail in \cref{sec:beam-test-setup}.
Similarly to the 2PA measurement, the pion beam was directed nearly perpendicular to the scintillator block, along the 5~cm short edge.

\begin{figure*}
	\centering
    \begin{subfigure}[b]{0.8\linewidth}\centering\includegraphics[width=\linewidth]{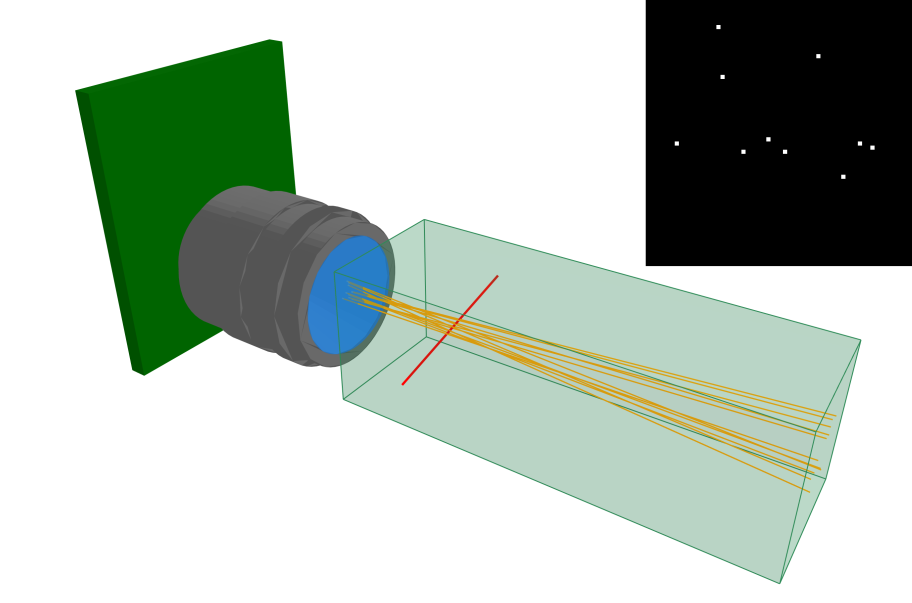}\caption{}\label{sfig:bt-event}\end{subfigure}
	\begin{subfigure}[t]{0.49\linewidth}\centering\includegraphics[width=\linewidth]{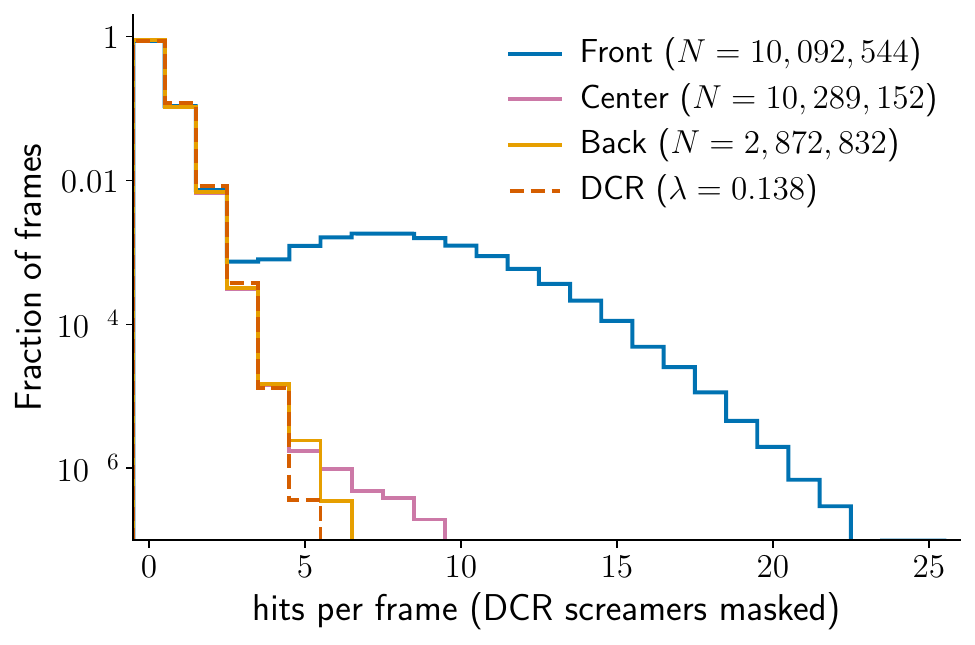}\caption{}\label{sfig:bt-hits}\end{subfigure}
    \begin{subfigure}[t]{0.49\linewidth}\centering\includegraphics[width=\linewidth]{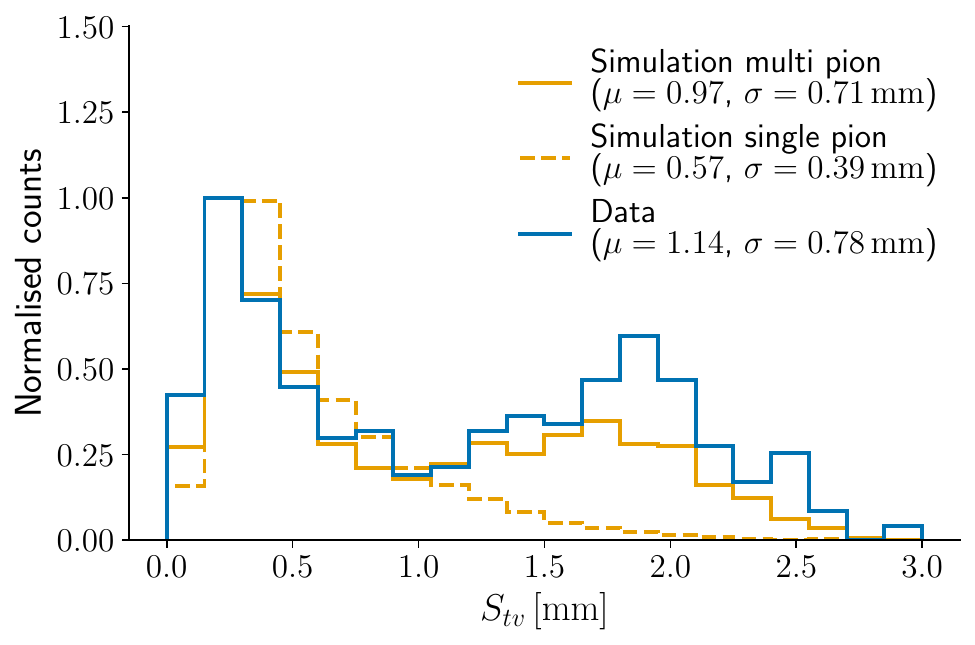}\caption{}\label{sfig:bt-sigma}\end{subfigure}
	\begin{subfigure}[t]{0.49\linewidth}\centering\includegraphics[width=\linewidth]{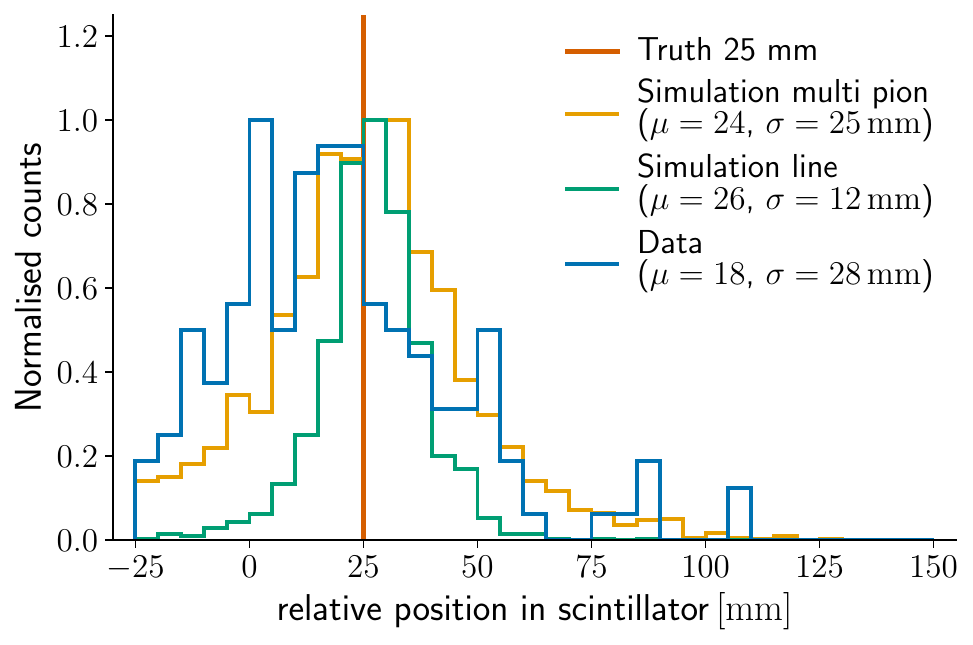}\caption{}\label{sfig:bt-depth}\end{subfigure}
	\begin{subfigure}[t] {0.49\linewidth}\centering\includegraphics[width=\linewidth]{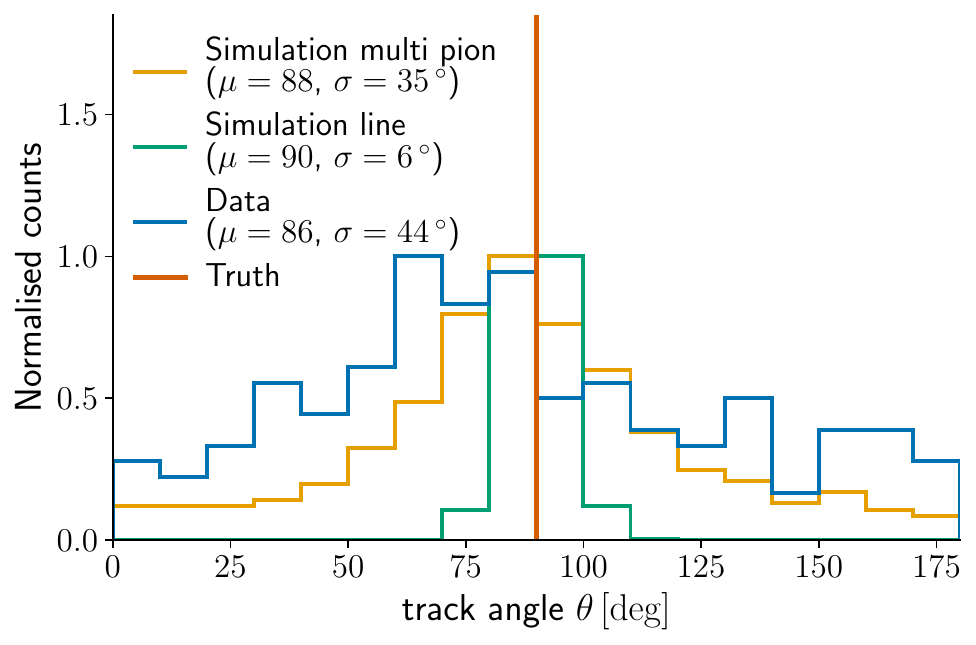}\caption{}\label{sfig:bt-angle}\end{subfigure}
	\caption{
    Results of the pion beam test data analysis.
    \subref{sfig:bt-event} Reconstructed 15 GeV/c pion data event.
    \subref{sfig:bt-hits} Number of counts per frame, without screamer pixels, for pions (solid) at the front (blue), central (green) and rear (orange) depth, compared to frames containing only dark counts (dashed red).
    \subref{sfig:bt-sigma} Transverse spread ($S_{tv}$) of the reconstructed position for pion data (blue), single-pion event simulation (green) and
	two-pion event simulation (orange).
    \subref{sfig:bt-depth} Reconstructed track depth in the scintillator for pion data (blue), pion simulation (orange), single-line track simulation (green) and
	a noise-only simulation (dashed black).
    The vertical dotted line at \SI{25}{\milli\metre} marks the true pion beam position.
    \subref{sfig:bt-angle} Reconstructed track angle ($\theta$) for pion data (blue) and simulation (orange).
    }
	\label{fig:beamtest}
\end{figure*}

The scintillator was exposed to the beam at three different depths: \SI{25}{\milli\metre} from its front face;
at \SI{75}{\milli\metre}, i.e. its centre;
and near its rear face.
The raw frames were processed exactly as in \cref{sec:2pa-reco-photon-starved}.
Pixels behind dead parts of the MLA (see \cref{sec:prototype}) were not used in the image post-processing.
After masking the \SI{5}{\percent} noisiest pixels, the residual dark count rate was \SI{0.14}{counts\per frame} over the full sensor, with at most $\sim 5$ dark counts in a single frame.
The light yield produced by pions in scintillator and detected by SwissSPAD2 is lower than that generated with 2PA laser.
Thus, the PLATON camera was moved closer to the scintillator block than during the 2PA measurements, as shown in \cref{sfig:bt-event}, but away from the best focus region of the main lens.
As a consequence, the larger solid angle subtended by the main lens increases the number of detected photons.
Moreover, a source nearer to the main lens illuminates a greater number of microlenses~\cite{Perwass2012}.
Hence, a wider portion of the solid angle can be sampled, providing better depth sensitivity.
On the other hand, this occurs at the cost of a reduced field of view and lateral resolution.
The resulting number of counts per frame is shown in \cref{sfig:bt-hits}.
The front sample shows a clear excess over the dark-count expectation, with a number of detected photons up to 20.
By contrast, the data samples acquired at depths farther from the camera are dominated by dark counts.
Thus, the image reconstruction was performed only for the data sample with the pion beam at the front of the scintillator.
Frames with at least \num{8} counts are selected and the same reconstruction method as described in \cref{sec:2pa-reco} and used for the 2PA data analysis is applied.

The reconstruction performance for test-beam pions is shown in \cref{fig:beamtest} and compared with our custom optical simulation validated in \cite{Dieminger2026} and \cref{sec:2pa-reco}.
The reconstructed depth, centred around the truth position, is shown in \cref{sfig:bt-depth}.
To better interpret the results, first, the optical simulation of a single-track source of uniformly distributed \numrange{8}{10} photons at the nominal pion beam position was performed using the same pipeline as for the 2PA simulations.
The large virtual depth yields a depth resolution down to about $\SI{12}{\milli\metre}$, confirming that depth reconstruction is feasible.
Then, we simulated the \SI{15}{\giga\electronvolt\per c} pion beam test.
The simulation predicts more counts than observed,
consistent with a realistic $\sim$\SIrange{15}{20}{\percent} light loss through the
main lens that is not accounted for in the optical model.
We therefore reweighted the simulated number of counts to match the data.

The transverse residual spread, $S_{tv}$ defined in \cref{eq:Stv}, of the fitted tracks (see \cref{sfig:bt-sigma}) shows two populations:
one is peaked below 0.5 mm and has a shape compatible with the simulation of single-pion events;
the other peaks around 2~mm.
The full distribution could be reproduced by simulating the beam pile-up, i.e. multiple pions per frame crossing the scintillator and distributed over the beam profile (\SI{16}{\milli\metre}).
While the distribution below 1~mm is dominated by frames in which only one pion supplied most of the photons
or the others fall outside the narrow field of view \SI{16}{\milli\metre},
the second peak is dominated by two pions equally contributing to the counted photons.
Therefore, to avoid overestimating the spatial resolution, we isolated single-pion events by selecting $S_{tv}<\SI{1}{\milli\metre}$.
Applying this selection to the front sample yields \num{157} events,
of which the reconstructed depth and angle $\theta$ (see definition in \cref{sfig:method-3d}) are shown, respectively, in \cref{sfig:bt-depth,sfig:bt-angle}.
Both agree relatively well with simulations.

The mean transverse spread of the selected events is \SI{0.41}{\milli\metre}, in agreement with the \SI{0.44}{\milli\metre} obtained from the pion simulation.
This is about twice the value measured for the 2PA tracks at a comparable number of detected photons (\cref{sfig:2pa-stv}).
The pion track can be considered genuinely thicker because it can include scattering or secondary particles created along its path in the scintillator.

The reconstructed depth, shown in \cref{sfig:bt-depth}, averages at \SI{18}{\milli\metre} from the scintillator front with a spread of \SI{28}{\milli\metre}, in relatively good agreement with the simulation, which has a mean reconstructed depth of \SI{24}{\milli\metre} and a spread of \SI{25}{\milli\metre}.
The remaining bias, still smaller than the spread, may be due to a potential misalignment of the pion beam with respect to its nominal position.
To confirm that this was not the result of random coincidence, we simulated and reconstructed \num{2000} events with only dark counts and no pion signal, each with an over-fluctuation of \numrange{8}{10} activated pixels.
One can see that the depth distribution of these events largely differs from the data.
Finally, relying on the reliability of the optical simulation, highlighted by the relatively good agreement with data, we extracted the prototype depth resolution from the reconstructed spread of simulated single-line light sources. This amounts to 12~mm.

As shown in \cref{sfig:bt-angle}, the average reconstructed angular spread is \SI{44}{\degree}.
A similar spread is observed also in the pion beam simulation.
It derives from the convolution of the detector angular resolution for photon-starved events and the contribution from pion scattering, which can undergo elastic or inelastic scattering, changing its direction and/or creating additional particles.
Similarly to the depth study, we can estimate the detector angular resolution by simulating a single line source without including effects of scattering physics processes or remnant pile-up. We obtain an angular resolution of \SI{6}{\degree}.

It is worth noting that, differently from the case of $\theta$, there is little resolution on the determination of the $\phi$ angle, for multiple reasons evident in \cref{fig:track-reco-method}.
First, the rotation plane is parallel to the camera optical axis and, thus, a change in $\phi$ is highly degenerate with the track depth, whose resolution is, by design, almost 10 times worse than that on the lateral position.
Second, this combines with the fact that, since the pion beam also lies in the $\phi$ rotation plane and crosses the scintillator block from left to right, a change in $\phi$ does not change the magnitude of the track length projected onto the photosensor.
Finally, the camera field of view allows only 16~mm of the track in the scintillator to be imaged, hence suffering from a relatively short lever arm.

Although the setup showed poor resolution in $\phi$, this issue could be easily overcome by a system of two orthogonal cameras, where the depth measurement of one camera corresponds to the lateral-position measurement of the other.
For instance, this would ensure a position spread of \SI{0.4}{\milli\metre}
for both measurements for 15 photons detected per camera,
as indicated in \cref{sfig:res-band}.

\section{Discussion}
\label{sec:discussions}

Our work paves the way towards high-resolution particle tracking in unsegmented scintillator with PLATON, a plenoptic camera instrumented with a SPAD array imaging sensor.
We demonstrated the feasibility of particle detection and track reconstruction with the PLATON prototype instrumented with SwissSPAD2.
After the characterisation with light tracks induced by laser two-photon absorption,
we succeeded in reconstructing 15 GeV/c pions produced at the CERN PS traversing a monolithic scintillator, finding good agreement with our custom optical simulation.

While a sub-millimetre lateral resolution is reached with only 15 detected photons, the depth resolution is about \SI{9.8}{\milli\metre} for 50 photons.
An improved spatial resolution, especially in depth, would require a photosensor with a higher light yield to reduce the effect of Poisson fluctuations in the photon-starved images, while preserving the time-resolving capabilities.
On the other hand, we note that in a setup with two orthogonally oriented PLATON cameras, a depth resolution below \SI{0.4}{\milli\metre}
could be achieved with only $\mathcal{O}(10)$ detected photons, as highlighted in \cref{sec:test-beam-reconstruction}, even with a beam not positioned on the optical focus.

It is worth noting that the PLATON prototype was operated at a numerical aperture of f/2.4. However, in another future prototype an optics with f/1 could be used owing to a different arrangement of the bonding wires and protective glob top, to allow a shorter sensor-to-MLA distance.
Moreover, issues during the assembly prevented the use of more than around 60\% of the active surface for the image post-processing (see \cref{sec:prototype} for details).
Hence, it is realistic to assume that in a future PLATON prototype, a light yield higher by a factor 10 is within reach without introducing major updates either to the optics or to the photosensor.
From this perspective, the depth and lateral resolutions for more than 100 detected photons could be taken as reference.

Among the possible future applications of PLATON, some would require particle tracking in large scintillator volumes, for example in a neutrino detector or in a homogeneous particle calorimeter.
In such cases, 3D track reconstruction in, for example, a $1 \times 1 \times 1$ m$^3$ scintillator surrounded by PLATON cameras would be further simplified, if $\mathcal{O}$(100 ps) single-photon timing information were available as well as a sufficient number of photons were detected.
In fact, while in a small scintillator block ambiguities arise from internal reflection or imperfect surface quality,
in a large volume these photons would be recorded with a measurable time delay, owing to the fast scintillator decay time.
Thus, it is interesting to note that the spatial resolution is not expected to degrade dramatically in a larger scintillator, as long as the volume is smaller than the photon scattering length, around \SI{3}{\metre} in PVT-based plastic scintillator~\cite{eljen-catalogue-2025} and around \SI{30}{\metre} in organic liquid scintillator \cite{ABUSLEME2021164823,10.1063/1.4927458}.

For a future large-scale PLATON detector, the development of SPAD array imaging sensors, or other photosensor technologies with analogous features, is key. The development of a SPAD array sensor with a photodetection efficiency closer to that of high-dynamic range silicon photomultipliers and single-photon sub-nanosecond time resolution is underway \cite{Kaneyasu2025PlatonSPAD}.

\section{Methods}

\subsection{PLATON prototype}
\label{sec:prototype}

The PLATON prototype, described in detail in ref.~\cite{Dieminger2026}, is a plenoptic camera built around a single-photon-sensitive imaging sensor to take fast 3D-photographs of charged particles in a \qtyproduct{50 x 50 x 150}{\milli\metre} block of EJ-262 organic scintillator \cite{datasheet_EJ_260_262}. For 1 MeV energy loss by an ionising particle, \num{8700} photons mainly in the blue range are produced. The attenuation length is \SI{250}{\centi\metre}, considerably longer than the size of the scintillator block. The rise time of \SI{0.9}{\nano\second} and the decay time of \SI{2.1}{\nano\second} ensure the emission of most of the scintillation photons within about \SI{10}{\nano\second}.

A main photographic lens collects light from the scene, which is then sent onto an MLA
and then onto the photosensor. The purpose of the MLA is to endow the detector with sensitivity to slightly different perspectives, so that the acquired light field encodes both the position and the direction of the incident light. This makes it possible to retrieve the depth of a source.

The sensor is a $512\times512$-pixel SwissSPAD2 gated \ac{spad} array~\cite{Ulku2019} with a \SI{16.38}{\micro\metre} pixel pitch and a peak photon-detection efficiency of \SI{5}{\percent} at \SI{520}{\nano\metre}.
Each frame is a binary single-photon image taken within a gate adjustable between \SI{10}{\nano\second} and \SI{100}{\micro\second}.
The Galilean focused-plenoptic-camera optics were produced and assembled by Raytrix GmbH \cite{raytrix} and operated at a numerical aperture of $f/2.4$.
A faster f-number, down to 1, could be achieved in another similar prototype with a different arrangement of the bonding wires and protective glob top, to allow for a shorter sensor-to-MLA distance.
During assembly, some of the glue used to fix the MLA leaked onto the sensor, leaving small regions on the left and right of the array without a functioning microlens. The corresponding pixels carry no usable light-field information and are excluded from the image reconstruction, thus reducing the effective light yield to about 60\% of the expected one.

\subsection{Two-photon absorption experimental setup}
\label{sec:setup}

\begin{figure*}[t]
	\centering
	\hfill\begin{subfigure}[b]{0.34\linewidth}\centering\includegraphics[width=\linewidth]{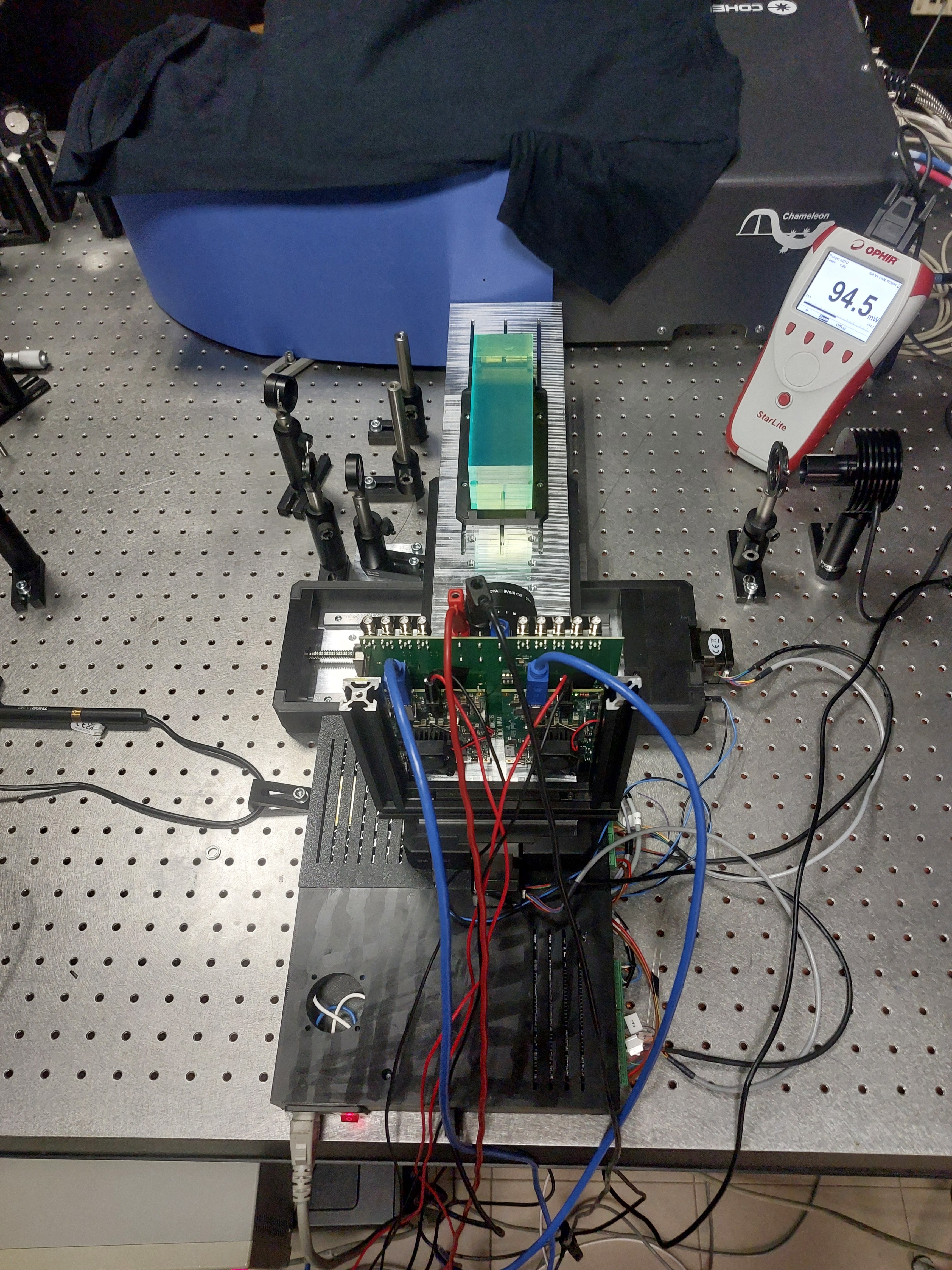}\caption{}\label{sfig:setup-table}\end{subfigure}\hfill
	\begin{subfigure}[b]{0.65\linewidth}\centering\includegraphics[width=\linewidth]{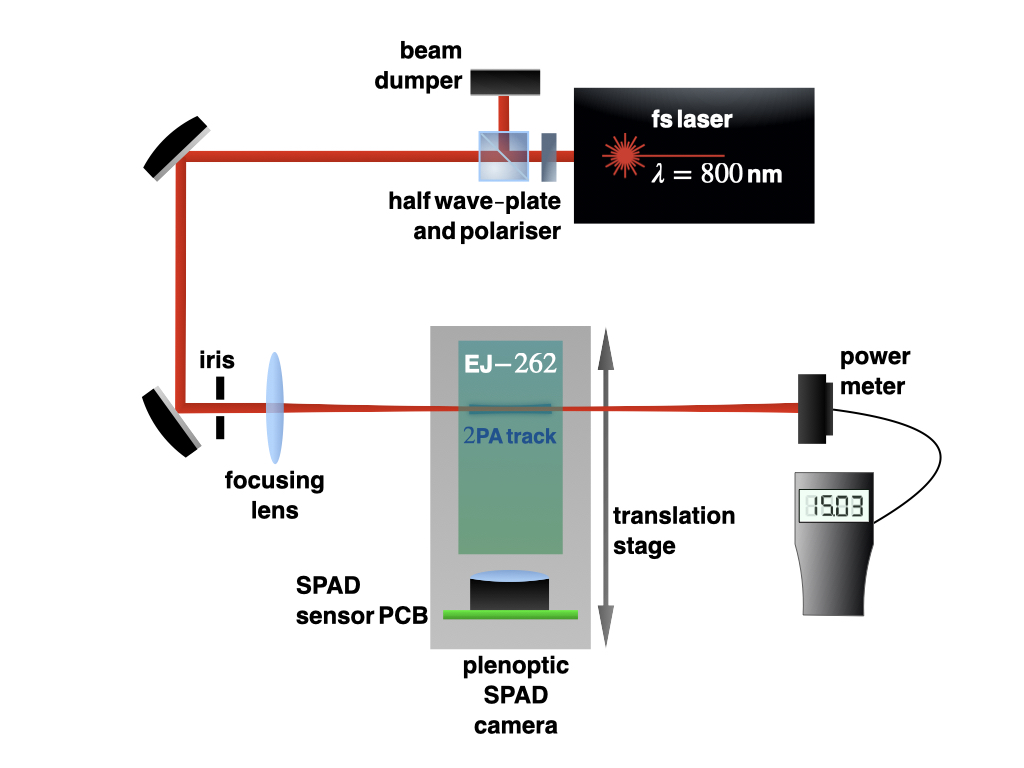}\caption{}\label{sfig:setup-sketch}\end{subfigure}\hfill

	\begin{subfigure}[t]{0.32\linewidth}\centering\includegraphics[width=\linewidth]{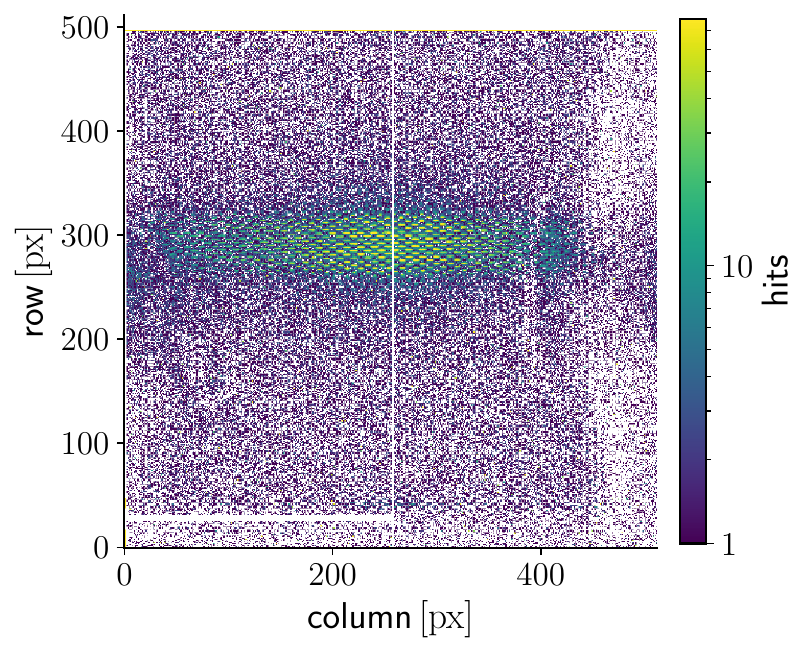}\caption{}\label{sfig:setup-summed-close}\end{subfigure}\hfill
	\begin{subfigure}[t]{0.32\linewidth}\centering\includegraphics[width=\linewidth]{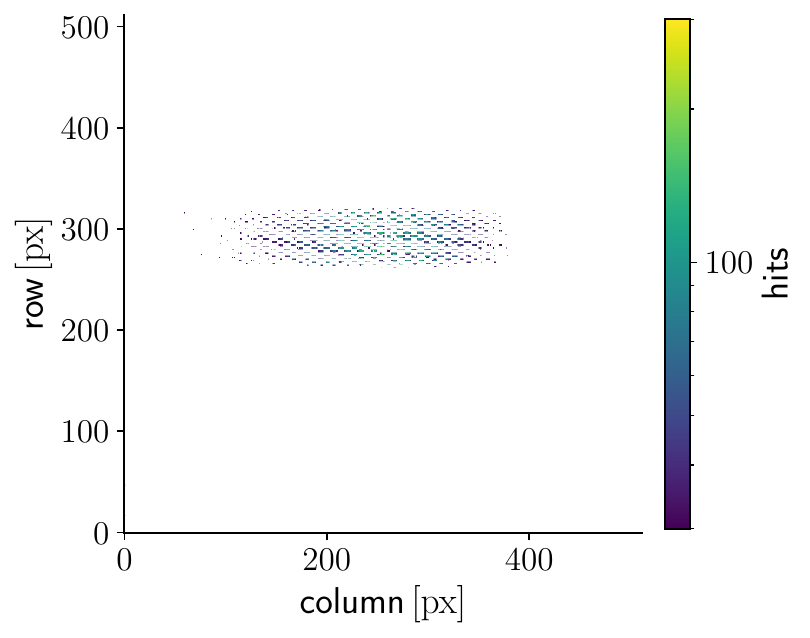}\caption{}\label{sfig:setup-thresh-close}\end{subfigure}\hfill
	\begin{subfigure}[t]{0.32\linewidth}\centering\includegraphics[width=0.83\linewidth]{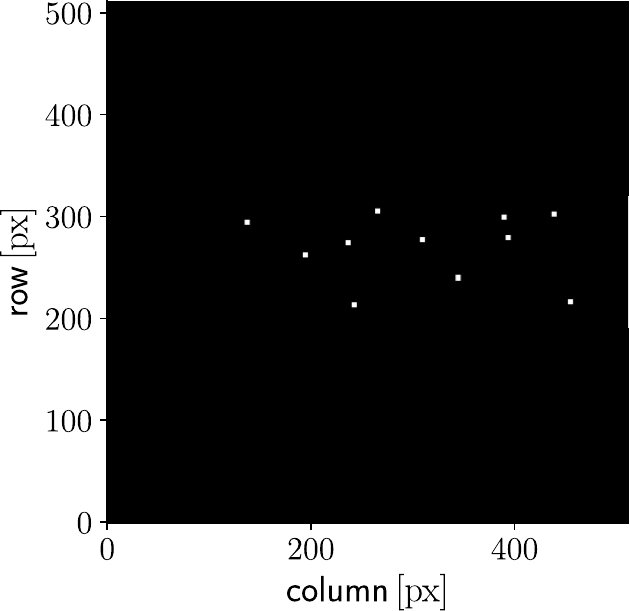}\caption{}\label{sfig:setup-frame-close}\end{subfigure}

	\begin{subfigure}[t]{0.32\linewidth}\centering\includegraphics[width=\linewidth]{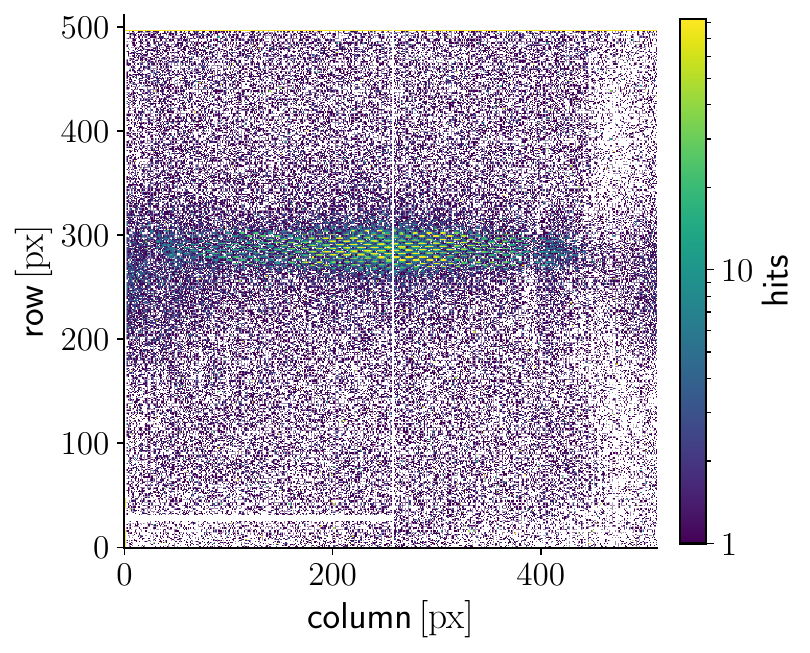}\caption{}\label{sfig:setup-summed-far}\end{subfigure}\hfill
	\begin{subfigure}[t]{0.32\linewidth}\centering\includegraphics[width=\linewidth]{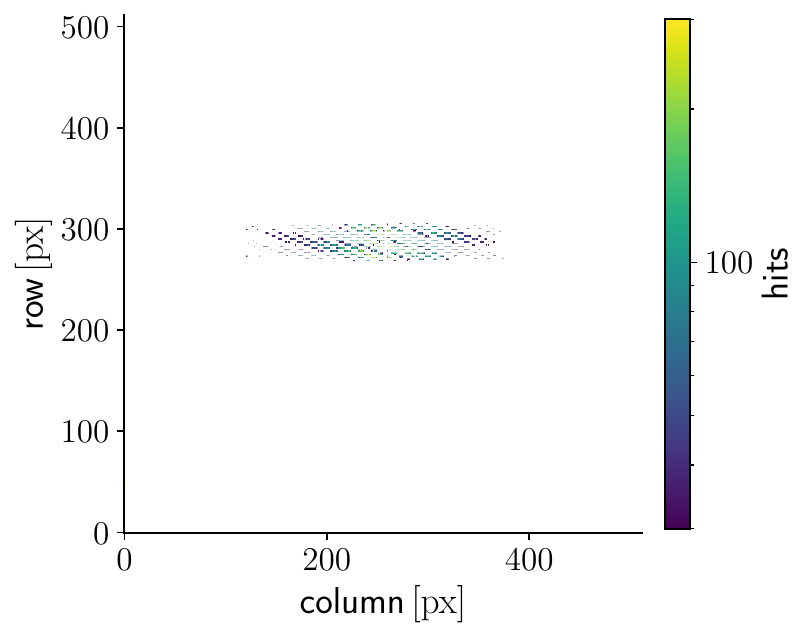}\caption{}\label{sfig:setup-thresh-far}\end{subfigure}\hfill
	\begin{subfigure}[t]{0.32\linewidth}\centering\includegraphics[width=0.83\linewidth]{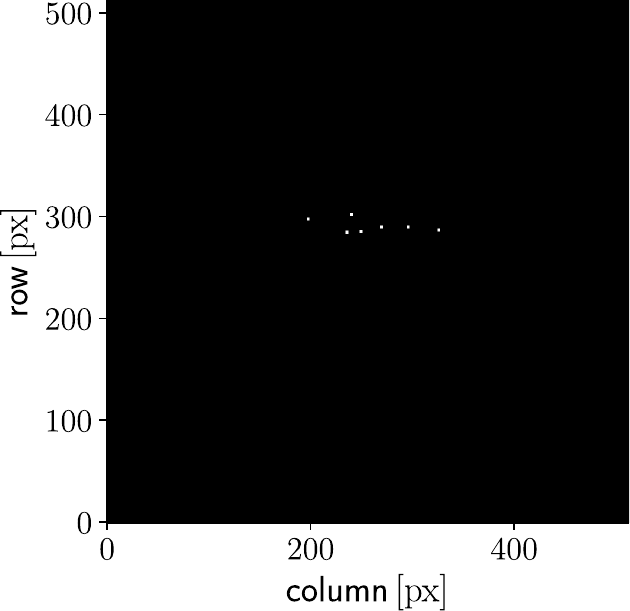}\caption{}\label{sfig:setup-frame-far}\end{subfigure}
	\caption{Experimental setup for the two-photon-absorption measurements.
	\subref{sfig:setup-table} The prototype on the optical table.
	\subref{sfig:setup-sketch} Sketch of the beam path: an \SI{800}{\nano\metre} femtosecond laser, attenuated by a half-wave plate and polariser (with the rejected beam sent to a dump), is focused by a lens into the EJ-262 scintillator mounted on the translation stage in front of the PLATON camera;
    the transmitted power is monitored behind the block.
    \subref{sfig:setup-summed-close}, \subref{sfig:setup-thresh-close}, \subref{sfig:setup-frame-close}
    show the track at the minimal scan depth (closest to the camera) and
    \subref{sfig:setup-summed-far}, \subref{sfig:setup-thresh-far}, \subref{sfig:setup-frame-far}
    at the maximal scan depth:
    the light intensity images
    \subref{sfig:setup-summed-close} and \subref{sfig:setup-summed-far},
    the same images after keeping only the brightest pixels
    \subref{sfig:setup-thresh-close} and \subref{sfig:setup-thresh-far},
    and two examples of single frame with $\mathcal{O}(10)$ activated pixels
    \subref{sfig:setup-frame-close} and \subref{sfig:setup-frame-far}.
    }
	\label{fig:setup}
	\label{fig:beampath}
    \label{fig:sumvsframe}\label{fig:summed_images}\label{fig:individual_frame}
\end{figure*}

The 2PA characterisation was carried out using a pulsed femtosecond infrared laser (model Coherent Chameleon Ultra, with average power \SI{2.5}{\watt}, repetition rate \SI{80}{\mega\hertz} and pulse duration \SI{171}{\femto\second} FWHM). The beam was delivered to the scintillator along the path sketched in Fig.~\ref{sfig:setup-sketch}.
A half-wave plate followed by a \ac{pbs} sets the pulse energy reaching the experiment, with the rejected polarisation directed into a beam dump. The transmitted beam is steered by two mirrors into a single \SI{50}{\milli\metre} focal-length lens, which focuses it into the scintillator block. The $1/e$ beam radius, measured in air, was \SI{26}{\micro\metre}.  The pulse energy was held fixed throughout the scan, and the power transmitted through the scintillator was continuously monitored by a power meter placed after it, allowing potential misalignments to be detected during the scan.
EJ-262 has an absorption band in the near-UV and emits over a broad band peaking around \SI{480}{\nano\metre}, with a fast decay time of \SI{2.7}{\nano\second}.
Linear absorption of infrared light, instead, is limited, thus granting the possibility of penetrating well within the scintillator. The simultaneous absorption of two photons at around \SI{800}{\nano\metre} is possible, due to the nonlinear response of EJ-262. Thus, such an excitation induces a re-emission similar to that produced by a traversing charged particle. The dependence of 2PA on the square of the intensity implies that this becomes significant only in the small, bright region around the focus~\cite{GoeppertMayer1931,KaiserGarrett1961,Denk1990}. Scintillation light produced in this region creates a straight narrow 2PA track oriented orthogonally to the optical axis of the camera, at the height of the optical axis.
The light is emitted isotropically, allowing the capability of the PLATON camera to acquire the light field independently of the light direction to be tested.
To align the laser beam along the sought directions, a 3D-printed cover with multiple aperture pairs at the height of the optical axis was mounted around the scintillator; see Fig.~\ref{fig:setup}. By maximising the power transmitted through each pair, we obtained a reliable reference ground truth for the position of the track -- their transverse \SI{2}{\milli\metre} size provides an indication on the tolerance. This cover was removed for data acquisition.

The beam was kept fixed, while different depths were explored by moving the camera and the scintillator along the optical axis by means of the motorised stage shown in \cref{fig:setup}. This allowed scans in the range \SIrange{200}{275}{\milli\metre} relative to the front of the lens, in steps of \SI{2.5}{\milli\metre} and with an accuracy better than \SI{100}{\micro\metre}. At every stage position, \num{163840} consecutive \SI{100}{\nano\second}-gated single-photon frames are captured off the SwissSPAD2 array, gated independently of the laser.

Isolation from stray light was achieved by shielding the whole setup by black cardboard in a fully darkened environment. Red light was blocked by a frequency-selective filter placed in front of the main lens, transmitting the blue scintillation light only. Measurements taken in the absence of any signal were used to verify the absence of significant ambient light and to identify noisy pixels.

\Cref{sfig:setup-summed-close} shows a raw image for a track close to the minimal achievable depth. One can recognise the pattern dictated by the disposition of the
MLA.
Upon closer inspection, differences due to the different viewpoint of each microlens can be appreciated. To filter the background noise, thresholding is applied, resulting in \cref{sfig:setup-thresh-close}.
In the photon-starved regime, typical raw images resemble that in \cref{sfig:setup-frame-close}, where it is evident that no sub-images can be identified, hence the need of the chief-ray method.
\Cref{sfig:setup-summed-far,sfig:setup-thresh-far,sfig:setup-frame-far} show similar results, this time for a farther track, near the maximal depth allowed by the scintillator size. These are broadly similar to the previous cases, but the variation in the conditions for collection appears clearly in the change in shape of the full image, as well as in the reduced brightness.

\subsection{Beam-test setup}
\label{sec:beam-test-setup}

\begin{figure*}
	\centering
	\begin{subfigure}[t]{0.37\linewidth}\centering\includegraphics[width=\linewidth,trim=302 52 804 83,clip]{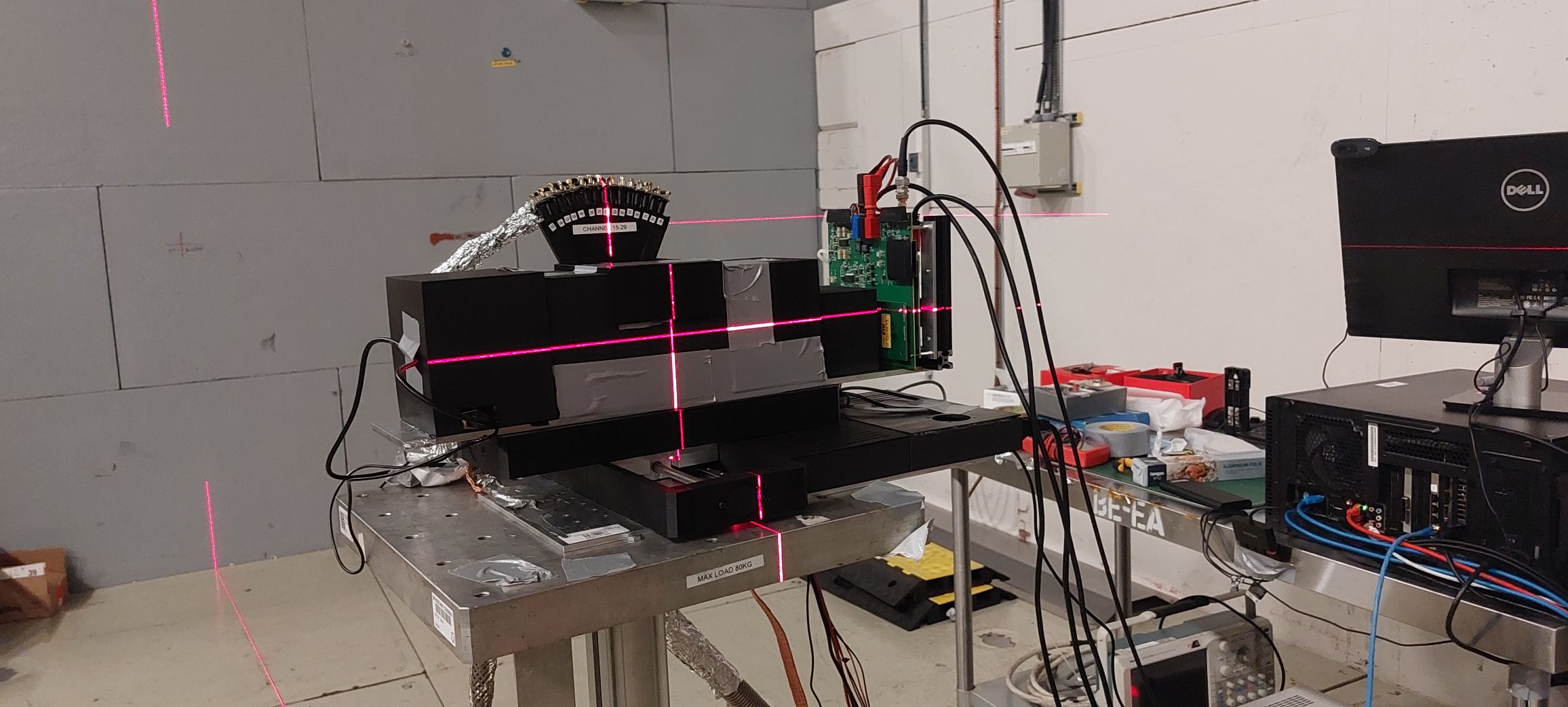}\caption{}\label{sfig:bt-photo-rail}\end{subfigure}
	\begin{subfigure}[t]{0.32\linewidth}\centering\includegraphics[width=\linewidth,trim=100 867 0 621,clip]{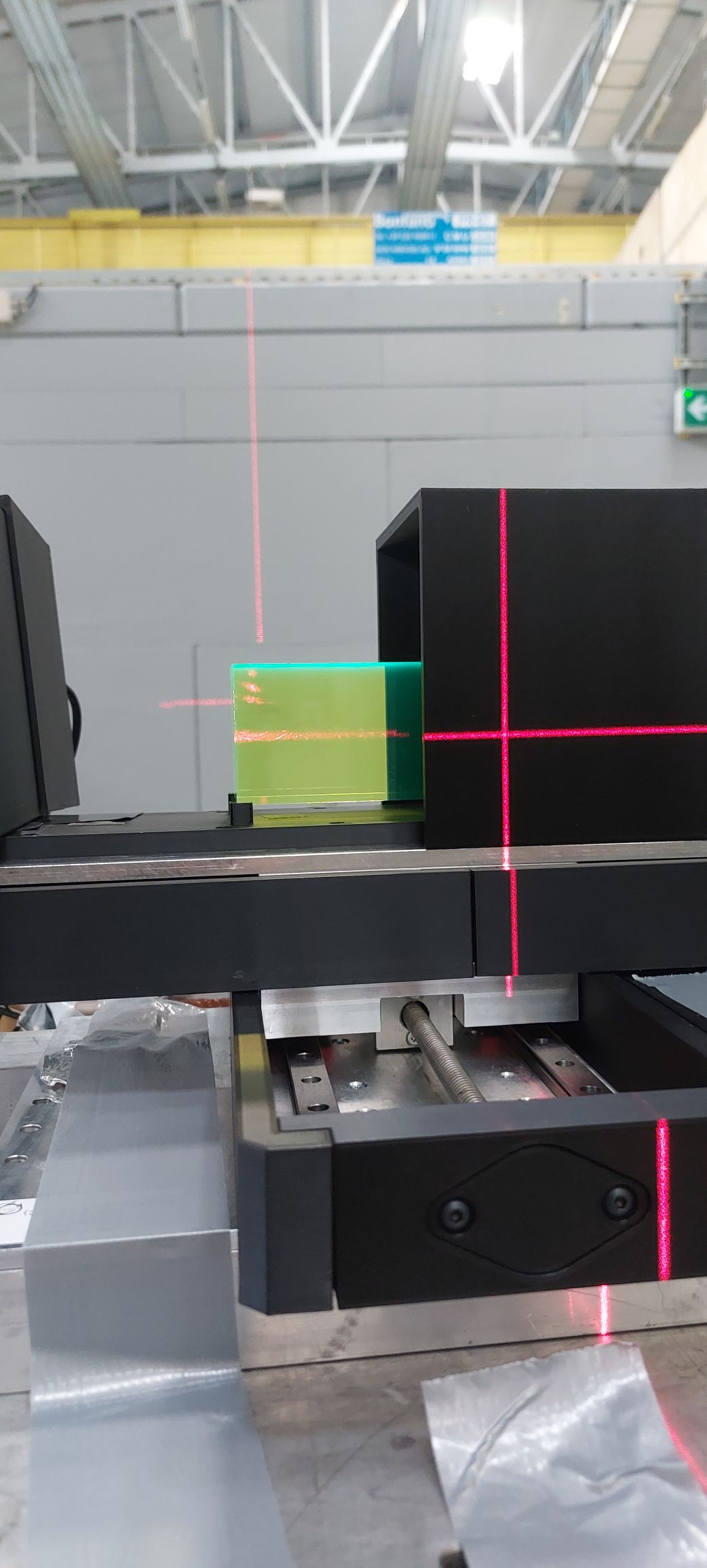}\caption{}\label{sfig:bt-photo-block}\end{subfigure}\hfill
	\begin{subfigure}[b]{\linewidth}
		\centering
		\includegraphics[width=0.32\linewidth]{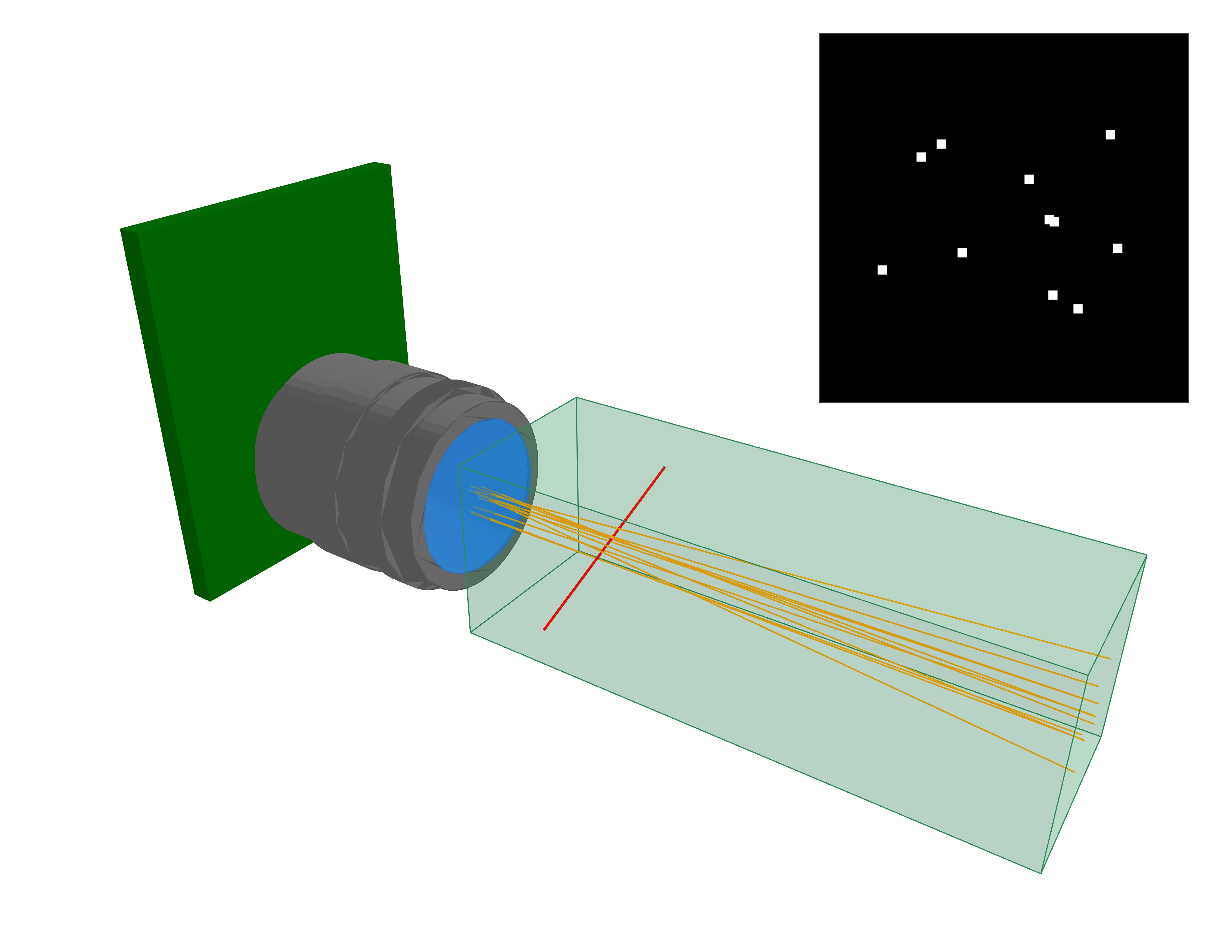}\hfill
		\includegraphics[width=0.32\linewidth]{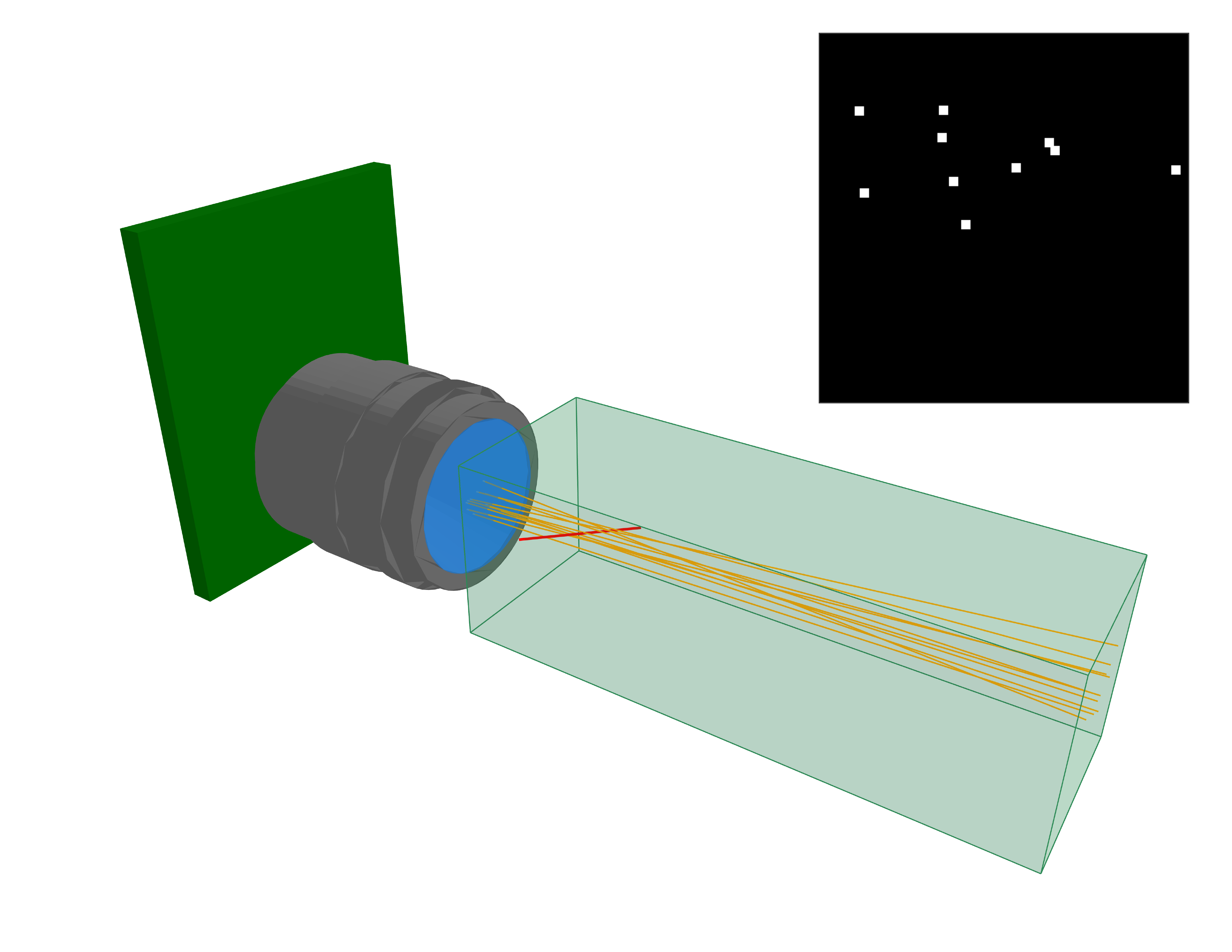}
		\includegraphics[width=0.32\linewidth]{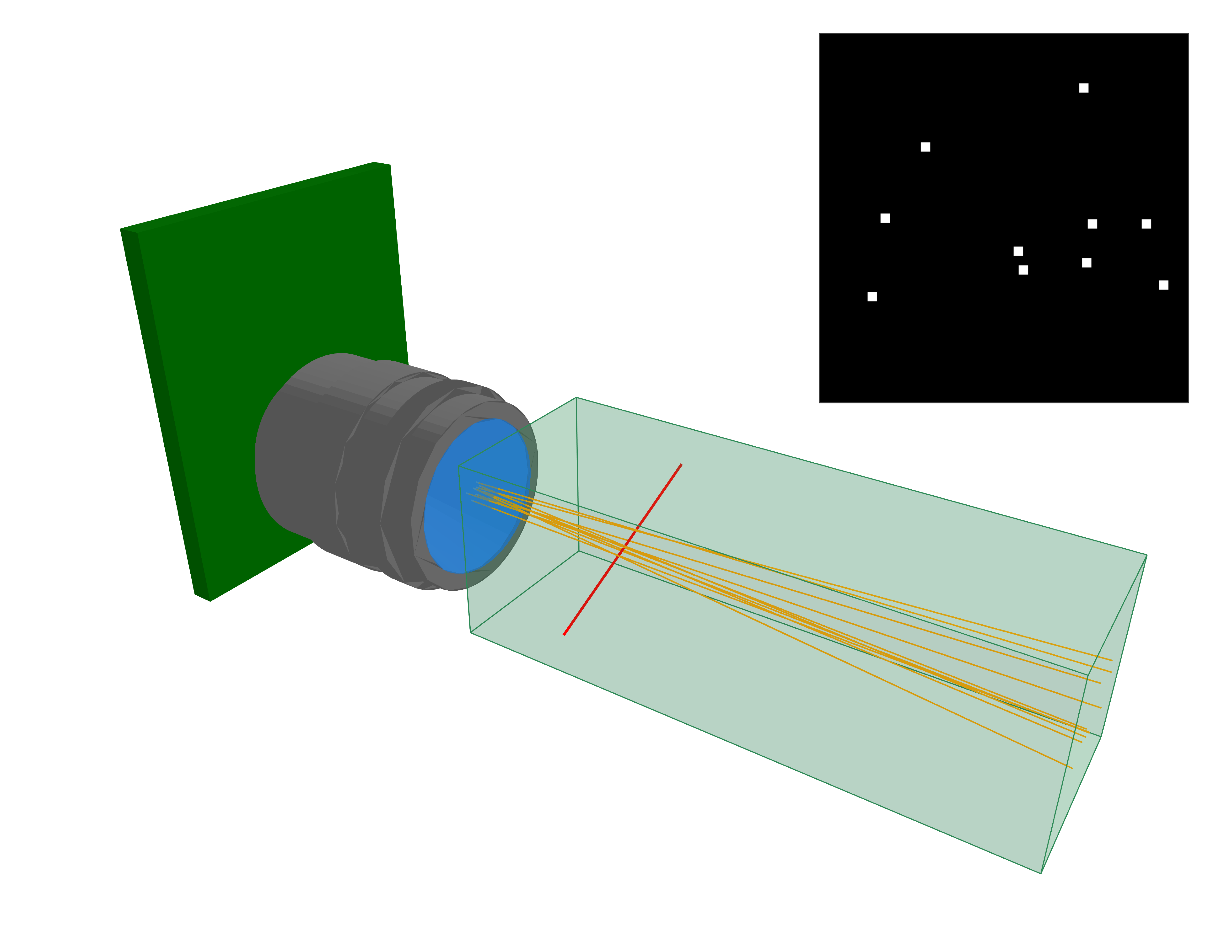}
	\caption{}\label{sfig:bt-frames}\end{subfigure}
	\caption{
    Setup and results of the beam test at the CERN PS beamline.
	\subref{sfig:bt-photo-rail},\subref{sfig:bt-photo-block} The prototype installed in the \SI{15}{\giga\electronvolt\per c}
	$\pi^-$ beam.
    \subref{sfig:bt-frames} Example frames of pion candidate signals. The white spots correspond to the activated pixels.
    The red lines show the fitted pion tracks.
    }
	\label{fig:beamtest-setup}
	\label{fig:beamtest-events}
\end{figure*}

For the beam test at CERN, the same equipment, as used for the 2PA measurement described in \cref{sec:setup}, was equipped with a 3D printed light-tight shroud, as shown in Figs.~\ref{sfig:bt-photo-rail} and \ref{sfig:bt-photo-block}, with the scintillator block now significantly closer to the sensor in order to increase the collection efficiency.
Under these conditions, the scintillator is out of focus, with its circle of minimal confusion larger than a pixel. Optical simulations indicate that sub-images have a point-spread function around \SI{40}{\micro\metre}.
A laser reference was used to align the camera and the scintillator with the pion beam (see \cref{sfig:bt-photo-rail,sfig:bt-photo-block}). The alignment was also verified by a two-layer scintillating fibre beam monitor connected to silicon photomultipliers.
Individual frames were collected during consecutive beam spills using a \SI{33}{\nano\second} exposure time, triggered by a signal available from the
beam pit. The scintillator was irradiated at three different locations over a range of \SI{100}{\milli\metre}: \SI{25}{\milli\metre} from the front of the scintillator (front position); in the centre of the scintillator (mid position); \SI{25}{\milli\metre} from the back of the scintillator (rear position). For the front and mid positions 10.1 million
frames were acquired, while for the rear positions 2.8 million frames were taken. The noise floor and the noisy pixels were identified in separate runs.
Examples of the collected tracks are reported in \cref{sfig:bt-frames}, showing the pixels activated by the scintillation light; these demonstrate the photon-starved conditions of typical experiments, and the relevance of the 2PA calibration.

\subsection{Reconstruction method}
\label{sec:fitting}

The retrieved chief rays are employed for track reconstruction. As introduced in the main text, each ray is reconstructed by assuming that a detected photon followed a trajectory from the scintillator through the optical system, identified as the line connecting the centre of the microlens with the centre of the activated pixel. The back-propagation algorithm models the main lens as a thick object, described by its principal planes $H_1$, towards the detector array, and $H_2$, towards the scintillator. A ray intersecting $H_2$ with an angle $\alpha_2$ and a height $h$ with respect to the optical axis is refracted by the lens and leaves $H_1$ with the new angle
\begin{equation}
	\alpha_{1} = \alpha_{2} - \frac{h}{f}.
      \label{eq:thicklens}
\end{equation}
The focal length and principal-plane positions are taken from a calibration of the camera~\cite{Dieminger2026}. The direction inside the scintillator volume accounts for Snell's law
at the air-scintillator interface.
Finally, the direction of each ray in the scintillator volume is obtained and can be used to determine the origin along the scintillation light track.

Our reference frame has its origin on the $H_2$ principal plane of the main lens, as shown in \cref{fig:track-reco-method}, with the $x$ direction aligned along the optical axis, while $y$ and $z$ are, respectively, the vertical and horizontal orthogonal axes. We denote the transverse plane at depth $d$ as $\Pi_d \{x=d\}$. The angles defining a track are thus the azimuthal angle $\theta$, formed with the $y$ direction in the $y$--$z$ plane, and the polar angle $\phi$, formed with the $x$ direction in the $x-z$ plane.
A track with $\theta = \phi = 0$ is a line lying in $\Pi_d$ parallel to $z$.
The rays intersect the candidate plane specified by depth $d$ at $N$ points with $y$-coordinates $y_i(d,\theta,\phi)$, $i = 1,\dots,N$, and tilted by $\phi$.
Their transverse spread is captured by
  \begin{equation}
  S_{tv}(d,\theta,\phi) =
\left[\frac{1}{N}\sum_{i=1}^{N}\big(y_i(d,\theta,\phi)-\bar{y}\big)^2\right]^{1/2},
  \label{eq:Stv}
  \end{equation}
which is minimal when the values $(d,\theta,\phi)$ match the actual track and grows as the plane moves away from the light source. Thus, we obtain the best-fit track as
\begin{equation}
  (\hat{d},\hat{\theta},\hat{\phi}) = \argmin_{d,\theta,\phi}\,S_{tv}(d,\theta,\phi),
\label{eq:minstd2d}
\end{equation}
using either gradient descent or, when unstable, a brute-force parameter scan.
The depth resolution with the typical number of detected photons is of the order of \SI{20}{\milli\metre} (see \cref{sec:2pa-reco-photon-starved}), which results in rather limited sensitivity in the reconstruction of $\phi$, also considering that the length of the track reconstructed in the photon-starved regime exceeds the size of the photosensor (see discussion in \cref{sec:test-beam-reconstruction}).
Therefore, we opted to keep $\phi=0$ in the fit \eqref{eq:minstd2d}.

Photosensors are affected by dark counts, which in the chief-ray method generates spurious rays along directions unrelated to the actual track. These produce an increase in the estimator \eqref{eq:minstd2d}, which can be significant in the photon-starved regime, and bias the retrieved depth and orientation to a spurious minimum. The fit itself can be used to curtail this effect by applying random sample consensus (RANSAC)~\cite{FischlerBolles:1981}. This procedure draws a random subset of rays, computes the estimator \eqref{eq:minstd2d}, and counts the number of rays whose closest approach to the track candidate is below $t=\SI{4}{\milli\metre}$ in Euclidean distance.
This assigns a rank to the specific ray configuration. Dark counts would artificially lower the rank, due to the possible presence of outliers in the distribution. Therefore, the candidate track is selected as the one with higher rank among 100 repetitions, {\it i.e.} as the one affected the least by the presence of stray rays - for equal rank, the track with the smallest spatial dispersion $S_{tv}$ is chosen.

\bmhead{Data availability}
Data underlying the results presented in this paper are not publicly
available at this time but may be obtained from the authors upon reasonable
request.

\bmhead{Code availability}
The code used in this work is not publicly available at this time due to ongoing intellectual property protection efforts. It may be made available upon request, subject to appropriate terms and, where applicable, the execution of a non-disclosure agreement.
\backmatter

\bmhead{Acknowledgements}
We thank the NA64 collaboration for offering us valuable beam time during their measurement run.
T.D. thanks Dr. Umut Kose and Dr. Katharina Lachner
for their support during the data acquisition of the beam-test data and their continuous availability for discussions.
We thank Mylenne Manrique and Gabriele Bizzarri for introducing T.D. to the details of the \acs{2pa} measurements, and Ilaria Gianani for discussion.
We would like to thank Prof.\ Andr\'e Rubbia from ETH Z\"urich for providing access to laboratory equipment and facilities.
This work was supported by the Swiss National Science Foundation
under grant PCEFP2\textunderscore203261. This research was also
partially supported by the Swiss National Science Foundation (grant 20QT21\textunderscore187716 Qu3D ``Quantum 3D Imaging at high speed and high resolution").
M.B.\
acknowledges support from MUR Dipartimento di Eccellenza 2023-2027, and
the NQSTI (CUP: B53C22004170006).

\bmhead{Author contributions}
D.S. conceived the PLATON detector, is the PI of the project funded by the Swiss National Science Foundation, and supervised every aspect of the project.
The plenoptic system of the prototype was designed and built by Raytrix GmbH.
E.C., C.B., P.M. and K.K. provided the SPAD array photosensor, assisted with its use, offered guidance, and supervised the project to ensure understanding of the results.
J.W. formulated the idea of using laser two-photon absorption measurements to characterise the prototype.
T.D. and T.W. prepared the setup and movement stage.
V.B., M.B. and T.D. ran the laser two-photon absorption experiment.
T.D. tested the prototype in the test beam at the CERN PS.
T.D. developed the simulation and image reconstruction software and analysed the data.
D.S., M.B, and V.B. supervised the data analysis of the 2PA data.
All authors contributed to the writing of the paper.

\bmhead{Competing interests}
The authors declare the following competing interests: T.D., S.A.-M., and D.S. are named inventors on a patent application filed by ETH Zurich related to the technology described in this article (status: pending). E.C. is a co-founder of NovoViz. NovoViz was not involved in this work or in the drafting of this paper. The other authors declare no competing interests.

\bibliography{biblio}

\begin{thebibliography}{10}
\expandafter\ifx\csname url\endcsname\relax
  \def\url#1{\burl{#1}}\fi
\expandafter\ifx\csname urlprefix\endcsname\relax\def\urlprefix{URL }\fi
\providecommand{\bibinfo}[2]{#2}
\providecommand{\eprint}[2][]{\url{#2}}
\providecommand{\doi}[1]{\url{https://doi.org/#1}}
\bibcommenthead

\bibitem{Kolanoski:2020ksk}
\bibinfo{author}{Kolanoski, H.} \& \bibinfo{author}{Wermes, N.}
\newblock \emph{\bibinfo{title}{Particle Detectors}}
  (\bibinfo{publisher}{Oxford University Press}, \bibinfo{address}{Oxford, UK},
  \bibinfo{year}{2020}).

\bibitem{Rubbia_2022}
\bibinfo{author}{Rubbia, A.}
\newblock \emph{\bibinfo{title}{Phenomenology of Particle Physics}}
  (\bibinfo{publisher}{Cambridge University Press}, \bibinfo{year}{2022}).

\bibitem{Hyper-Kamiokande:2025fci}
\bibinfo{author}{Abe, K.} \emph{et~al.}
\newblock \bibinfo{title}{{Sensitivity of the Hyper-Kamiokande experiment to
  neutrino oscillation parameters using accelerator neutrinos}}.
\newblock \emph{\bibinfo{journal}{Eur. Phys. J. C}}
  \textbf{\bibinfo{volume}{86}} (\bibinfo{year}{2026}).

\bibitem{Hyper-Kamiokande:2018ofw}
\bibinfo{author}{Abe, K.} \emph{et~al.}
\newblock \bibinfo{title}{{Hyper-Kamiokande Design Report}}.
\newblock \bibinfo{type}{Design Report}, \bibinfo{institution}{Hyper-Kamiokande
  Collaboration} (\bibinfo{year}{2018}).
\newblock \bibinfo{note}{ArXiv:1805.04163},
  \bibinfo{eprint}{{\href{https://arxiv.org/abs/1805.04163}{{arXiv:1805.04163}}}}.

\bibitem{Abi:2020qib}
\bibinfo{author}{Abi, B.} \emph{et~al.}
\newblock \bibinfo{title}{{Long-baseline neutrino oscillation physics potential
  of the DUNE experiment}}.
\newblock \emph{\bibinfo{journal}{Eur. Phys. J. C}}
  \textbf{\bibinfo{volume}{80}}, \bibinfo{pages}{978} (\bibinfo{year}{2020}).

\bibitem{Abe:2019vii}
\bibinfo{author}{Abe, K.} \emph{et~al.}
\newblock \bibinfo{title}{{Constraint on the matter--antimatter
  symmetry-violating phase in neutrino oscillations}}.
\newblock \emph{\bibinfo{journal}{Nature}} \textbf{\bibinfo{volume}{580}},
  \bibinfo{pages}{339--344} (\bibinfo{year}{2020}).
\newblock \bibinfo{note}{[Erratum: Nature 583, E16 (2020)]}.

\bibitem{NOvA:2022see}
\bibinfo{author}{Acero, M.~A.} \emph{et~al.}
\newblock \bibinfo{title}{Measurement of the ${\ensuremath{\nu}}_{e}$-nucleus
  charged-current double-differential cross section at
  $\langle{E}_{\ensuremath{\nu}}\rangle=2.4\,\mathrm{GeV}$ using nova}.
\newblock \emph{\bibinfo{journal}{Phys. Rev. Lett.}}
  \textbf{\bibinfo{volume}{130}}, \bibinfo{pages}{051802}
  (\bibinfo{year}{2023}).
\newblock
  \urlprefix\url{https://link.aps.org/doi/10.1103/PhysRevLett.130.051802}.

\bibitem{MINOS:2008hdf}
\bibinfo{author}{Michael, D.~G.} \emph{et~al.}
\newblock \bibinfo{title}{{The Magnetized steel and scintillator calorimeters
  of the MINOS experiment}}.
\newblock \emph{\bibinfo{journal}{Nucl. Instrum. Meth. A}}
  \textbf{\bibinfo{volume}{596}}, \bibinfo{pages}{190--228}
  (\bibinfo{year}{2008}).

\bibitem{dama-libra}
\bibinfo{author}{Bernabei, R.} \emph{et~al.}
\newblock \bibinfo{title}{Final model independent result of
  dama/libra--phase1}.
\newblock \emph{\bibinfo{journal}{The European Physical Journal C}}
  \textbf{\bibinfo{volume}{73}}, \bibinfo{pages}{2648} (\bibinfo{year}{2013}).
\newblock \urlprefix\url{https://doi.org/10.1140/epjc/s10052-013-2648-7}.

\bibitem{doi:10.1126/sciadv.adv6503}
\bibinfo{author}{Carlin, N.} \emph{et~al.}
\newblock \bibinfo{title}{Cosine-100 full dataset challenges the annual
  modulation signal of dama/libra}.
\newblock \emph{\bibinfo{journal}{Science Advances}}
  \textbf{\bibinfo{volume}{11}}, \bibinfo{pages}{eadv6503}
  (\bibinfo{year}{2025}).
\newblock
  \urlprefix\url{https://www.science.org/doi/abs/10.1126/sciadv.adv6503}.

\bibitem{ANDREEV2026170830}
\bibinfo{author}{Andreev, Y.~M.} \emph{et~al.}
\newblock \bibinfo{title}{High efficiency veto hadron calorimeter in the na64
  experiment at cern}.
\newblock \emph{\bibinfo{journal}{Nuclear Instruments and Methods in Physics
  Research Section A: Accelerators, Spectrometers, Detectors and Associated
  Equipment}} \textbf{\bibinfo{volume}{1081}}, \bibinfo{pages}{170830}
  (\bibinfo{year}{2026}).
\newblock
  \urlprefix\url{https://www.sciencedirect.com/science/article/pii/S0168900225006321}.

\bibitem{DarkSide-20k:2024yfq}
\bibinfo{author}{Acerbi, F.} \emph{et~al.}
\newblock \bibinfo{title}{{DarkSide-20k sensitivity to light dark matter
  particles}}.
\newblock \emph{\bibinfo{journal}{Commun. Phys.}} \textbf{\bibinfo{volume}{7}},
  \bibinfo{pages}{422} (\bibinfo{year}{2024}).

\bibitem{Aprile:2020vtw}
\bibinfo{author}{Aprile, E.} \emph{et~al.}
\newblock \bibinfo{title}{{Projected WIMP sensitivity of the XENONnT dark
  matter experiment}}.
\newblock \emph{\bibinfo{journal}{JCAP}} \textbf{\bibinfo{volume}{11}},
  \bibinfo{pages}{031} (\bibinfo{year}{2020}).

\bibitem{CMSHCAL:2007zcq}
\bibinfo{author}{Abdullin, S.} \emph{et~al.}
\newblock \bibinfo{title}{{Design, performance, and calibration of CMS
  hadron-barrel calorimeter wedges}}.
\newblock \emph{\bibinfo{journal}{Eur. Phys. J. C}}
  \textbf{\bibinfo{volume}{55}}, \bibinfo{pages}{159--171}
  (\bibinfo{year}{2008}).

\bibitem{CMS:2009cmd}
\bibinfo{author}{Chatrchyan, S.} \emph{et~al.}
\newblock \bibinfo{title}{{Performance and Operation of the CMS Electromagnetic
  Calorimeter}}.
\newblock \emph{\bibinfo{journal}{JINST}} \textbf{\bibinfo{volume}{5}},
  \bibinfo{pages}{T03010} (\bibinfo{year}{2010}).

\bibitem{Sgalaberna:2017khy}
\bibinfo{author}{Blondel, A.} \emph{et~al.}
\newblock \bibinfo{title}{A fully-active fine-grained detector with three
  readout views}.
\newblock \emph{\bibinfo{journal}{Journal of Instrumentation}}
  \textbf{\bibinfo{volume}{13}}, \bibinfo{pages}{P02006--P02006}
  (\bibinfo{year}{2018}).
\newblock
  \urlprefix\url{https://doi.org/10.1088\%2F1748-0221\%2F13\%2F02\%2Fp02006}.

\bibitem{T2K:2026zms}
\bibinfo{author}{Abe, S.} \emph{et~al.}
\newblock \bibinfo{title}{{The super fine-grained detector for the T2K neutrino
  oscillation experiment}}.
\newblock \emph{\bibinfo{journal}{Nucl. Instrum. Meth. A}}
  \textbf{\bibinfo{volume}{1092}}, \bibinfo{pages}{171882}
  (\bibinfo{year}{2026}).

\bibitem{Joram:2015ymp}
\bibinfo{author}{Joram, C.} \emph{et~al.}
\newblock \bibinfo{title}{{LHCb Scintillating Fibre Tracker Engineering Design
  Review Report: Fibres, Mats and Modules}}.
\newblock \bibinfo{type}{Tech. Rep.}, \bibinfo{institution}{CERN}
  (\bibinfo{year}{2015}).

\bibitem{Papa:2023uqv}
\bibinfo{author}{Papa, A.} \emph{et~al.}
\newblock \bibinfo{title}{{The Mu3e scintillating fiber detector R\&D}}.
\newblock \emph{\bibinfo{journal}{Nucl. Instrum. Meth. A}}
  \textbf{\bibinfo{volume}{1050}}, \bibinfo{pages}{168099}
  (\bibinfo{year}{2023}).

\bibitem{Dieminger2026}
\bibinfo{author}{Dieminger, T.} \emph{et~al.}
\newblock \bibinfo{title}{An ultrafast plenoptic-camera system for
  high-resolution 3d particle tracking in unsegmented scintillators}.
\newblock \emph{\bibinfo{journal}{Nature Communications}}
  \textbf{\bibinfo{volume}{17}}, \bibinfo{pages}{4204} (\bibinfo{year}{2026}).

\bibitem{Lippmann_1908_Epreuves_reversibles_donnant}
\bibinfo{author}{Lippmann, G.}
\newblock \bibinfo{title}{{\'E}preuves r{\'e}versibles donnant la sensation du
  relief}.
\newblock \emph{\bibinfo{journal}{Journal de Physique Th{\'e}orique et
  Appliqu{\'e}e}} \textbf{\bibinfo{volume}{7}}, \bibinfo{pages}{821--825}
  (\bibinfo{year}{1908}).
\newblock \urlprefix\url{http://dx.doi.org/10.1051/jphystap:019080070082100}.

\bibitem{Adelson_1992_Single_lens_stereo}
\bibinfo{author}{Adelson, E.} \& \bibinfo{author}{Wang, J.}
\newblock \bibinfo{title}{Single lens stereo with a plenoptic camera}.
\newblock \emph{\bibinfo{journal}{IEEE Transactions on Pattern Analysis and
  Machine Intelligence}} \textbf{\bibinfo{volume}{14}},
  \bibinfo{pages}{99--106} (\bibinfo{year}{1992}).

\bibitem{2005_Ng_Handheld_lightfield}
\bibinfo{author}{Ng, R.} \emph{et~al.}
\newblock \bibinfo{title}{Light field photography with a hand-held plenoptic
  camera}.
\newblock \emph{\bibinfo{journal}{CTSR}}  (\bibinfo{year}{2005}).
\newblock \urlprefix\url{https://graphics.stanford.edu/papers/lfcamera/}.

\bibitem{2006_Ng_Digital_Lightfield_Photography}
\bibinfo{author}{Ng, R.}
\newblock \emph{\bibinfo{title}{Digital Light Field Photography}}.
\newblock Ph.D. thesis, \bibinfo{school}{Department of Computer Science}
  (\bibinfo{year}{2006}).
\newblock
  \urlprefix\url{https://people.eecs.berkeley.edu/\~ren/thesis/renng-thesis.pdf}.

\bibitem{Levoy_1996_Light_field_rendering}
\bibinfo{author}{Levoy, M.} \& \bibinfo{author}{Hanrahan, P.}
  \emph{\bibinfo{title}{Light field rendering}}.
\newblock \emph{\bibinfo{booktitle}{Proceedings of the 23rd Annual Conference
  on Computer Graphics and Interactive Techniques}}, SIGGRAPH '96,
  \bibinfo{pages}{31–42} (\bibinfo{publisher}{Association for Computing
  Machinery}, \bibinfo{address}{New York, NY, USA}, \bibinfo{year}{1996}).
\newblock \urlprefix\url{https://doi.org/10.1145/237170.237199}.

\bibitem{Ulku2019}
\bibinfo{author}{Ulku, A.~C.} \emph{et~al.}
\newblock \bibinfo{title}{A 512 \texttimes{} 512 spad image sensor with
  integrated gating for widefield flim}.
\newblock \emph{\bibinfo{journal}{IEEE Journal of Selected Topics in Quantum
  Electronics}} \textbf{\bibinfo{volume}{25}}, \bibinfo{pages}{1--12}
  (\bibinfo{year}{2019}).
\newblock \urlprefix\url{http://dx.doi.org/10.1109/jstqe.2018.2867439}.

\bibitem{GoeppertMayer1931}
\bibinfo{author}{G{\"o}ppert-Mayer, M.}
\newblock \bibinfo{title}{{\"U}ber elementarakte mit zwei quantenspr{\"u}ngen}.
\newblock \emph{\bibinfo{journal}{Ann. Phys.}} \textbf{\bibinfo{volume}{401}},
  \bibinfo{pages}{273--294} (\bibinfo{year}{1931}).

\bibitem{Denk1990}
\bibinfo{author}{Denk, W.}, \bibinfo{author}{Strickler, J.~H.} \&
  \bibinfo{author}{Webb, W.~W.}
\newblock \bibinfo{title}{Two-photon laser scanning fluorescence microscopy}.
\newblock \emph{\bibinfo{journal}{Science}} \textbf{\bibinfo{volume}{248}},
  \bibinfo{pages}{73--76} (\bibinfo{year}{1990}).

\bibitem{Auffray2015}
\bibinfo{author}{Auffray, E.} \emph{et~al.}
\newblock \bibinfo{title}{Application of two-photon absorption in {PWO}
  scintillator for fast timing of interaction with ionizing radiation}.
\newblock \emph{\bibinfo{journal}{Nucl. Instrum. Methods Phys. Res. A}}
  \textbf{\bibinfo{volume}{804}}, \bibinfo{pages}{194--200}
  (\bibinfo{year}{2015}).

\bibitem{Perwass2012}
\bibinfo{author}{Perwa{\ss}, C.} \& \bibinfo{author}{Wietzke, L.}
  \emph{\bibinfo{title}{Single lens 3{D}-camera with extended depth-of-field}}.
\newblock \emph{\bibinfo{booktitle}{Human Vision and Electronic Imaging XVII}},
  Vol. \bibinfo{volume}{8291} of \emph{\bibinfo{series}{Proc. SPIE}},
  \bibinfo{pages}{829108} (\bibinfo{year}{2012}).

\bibitem{Adelson1992}
\bibinfo{author}{Adelson, E.~H.} \& \bibinfo{author}{Wang, J. Y.~A.}
\newblock \bibinfo{title}{Single lens stereo with a plenoptic camera}.
\newblock \emph{\bibinfo{journal}{IEEE Trans. Pattern Anal. Mach. Intell.}}
  \textbf{\bibinfo{volume}{14}}, \bibinfo{pages}{99--106}
  (\bibinfo{year}{1992}).

\bibitem{Jeon2015}
\bibinfo{author}{Jeon, H.-G.} \emph{et~al.} \emph{\bibinfo{title}{Accurate
  depth map estimation from a lenslet light field camera}}.
\newblock \emph{\bibinfo{booktitle}{Proceedings of the IEEE Conference on
  Computer Vision and Pattern Recognition (CVPR)}}, \bibinfo{pages}{1547--1555}
  (\bibinfo{year}{2015}).

\bibitem{Ng2005}
\bibinfo{author}{Ng, R.} \emph{et~al.}
\newblock \bibinfo{title}{Light field photography with a hand-held plenoptic
  camera}.
\newblock \bibinfo{type}{Tech. Rep.} \bibinfo{number}{CSTR 2005-02},
  \bibinfo{institution}{Stanford University} (\bibinfo{year}{2005}).

\bibitem{datasheet_EJ_260_262}
\bibinfo{author}{Eljen}.
\newblock \emph{\bibinfo{title}{Green Emitting Plastic Scintillator EJ-260 and
  EJ-262}} (\bibinfo{year}{2016}).
\newblock
  \urlprefix\url{https://eljentechnology.com/products/plastic-scintillators/ej-260-ej-262}.

\bibitem{eljen-catalogue-2025}
\bibinfo{author}{Eljen}.
\newblock \emph{\bibinfo{title}{Organic Scintillators Product Catalog}}
  (\bibinfo{year}{2025}).
\newblock
  \urlprefix\url{https://eljentechnology.com/images/technical_library/Eljen-Catalog-2025-01-web.pdf}.

\bibitem{ABUSLEME2021164823}
\bibinfo{author}{Abusleme, A.} \emph{et~al.}
\newblock \bibinfo{title}{Optimization of the juno liquid scintillator
  composition using a daya bay antineutrino detector}.
\newblock \emph{\bibinfo{journal}{Nuclear Instruments and Methods in Physics
  Research Section A: Accelerators, Spectrometers, Detectors and Associated
  Equipment}} \textbf{\bibinfo{volume}{988}}, \bibinfo{pages}{164823}
  (\bibinfo{year}{2021}).
\newblock
  \urlprefix\url{https://www.sciencedirect.com/science/article/pii/S0168900220312201}.

\bibitem{10.1063/1.4927458}
\bibinfo{author}{Zhou, X.} \emph{et~al.}
\newblock \bibinfo{title}{Rayleigh scattering of linear alkylbenzene in large
  liquid scintillator detectors}.
\newblock \emph{\bibinfo{journal}{Review of Scientific Instruments}}
  \textbf{\bibinfo{volume}{86}}, \bibinfo{pages}{073310}
  (\bibinfo{year}{2015}).
\newblock \urlprefix\url{https://doi.org/10.1063/1.4927458}.

\bibitem{Kaneyasu2025PlatonSPAD}
\bibinfo{author}{Kaneyasu, K.} \emph{et~al.} \emph{\bibinfo{title}{{PlatonSPAD:
  A novel SPAD sensor for large-scale high-resolution particle detectors}}}.
\newblock \emph{\bibinfo{booktitle}{Proceedings of the International Image
  Sensor Workshop (IISW)}}, \bibinfo{pages}{1--4} (\bibinfo{year}{2025}).

\bibitem{raytrix}
\bibinfo{author}{{Raytrix GmbH}}.
\newblock \bibinfo{title}{Raytrix}.
\newblock \bibinfo{howpublished}{\url{https://raytrix.de}}.
\newblock \bibinfo{note}{Accessed: 2025-08-01}.

\bibitem{KaiserGarrett1961}
\bibinfo{author}{Kaiser, W.} \& \bibinfo{author}{Garrett, C. G.~B.}
\newblock \bibinfo{title}{Two-photon excitation in {CaF$_2$:Eu$^{2+}$}}.
\newblock \emph{\bibinfo{journal}{Phys. Rev. Lett.}}
  \textbf{\bibinfo{volume}{7}}, \bibinfo{pages}{229--231}
  (\bibinfo{year}{1961}).

\bibitem{FischlerBolles:1981}
\bibinfo{author}{Fischler, M.~A.} \& \bibinfo{author}{Bolles, R.~C.}
\newblock \bibinfo{title}{Random sample consensus: A paradigm for model fitting
  with applications to image analysis and automated cartography}.
\newblock \emph{\bibinfo{journal}{Communications of the ACM}}
  \textbf{\bibinfo{volume}{24}}, \bibinfo{pages}{381--395}
  (\bibinfo{year}{1981}).

\end{thebibliography}

\end{document}